\documentclass{article}

\usepackage[preprint]{neurips_2026}

\usepackage[utf8]{inputenc}
\usepackage[T1]{fontenc}
\usepackage{hyperref}
\usepackage{url}
\usepackage[most]{tcolorbox}

\definecolor{BoxLightBlue}{RGB}{230,242,255}
\definecolor{BoxDarkBlue}{RGB}{30,90,150}
\usepackage{booktabs}
\usepackage{amsfonts}
\usepackage{amsmath}
\usepackage{amssymb}
\usepackage{nicefrac}
\usepackage{microtype}
\usepackage{xcolor}
\usepackage{colortbl}
\usepackage{graphicx}
\usepackage{enumitem}
\usepackage{algorithm}
\usepackage{algpseudocode}
\usepackage{subcaption}
\usepackage{multirow}
\usepackage[normalem]{ulem}

\title{%
  SkillOps: Managing LLM Agent Skill Libraries as Self-Maintaining Software Ecosystems
}

\author{%
  Xinyuan Song \\
  Emory University \\
  \texttt{xsong69@emory.edu}
  \And
  Hongji Pu \\
  University of Illinois Urbana-Champaign \\
  \texttt{hongjip2@illinois.edu}
  \And
  Liang Zhao\thanks{Corresponding author.} \\
  Emory University \\
  \texttt{liang.zhao@emory.edu}
}

\begin{document}

\maketitle
\begin{abstract}
LLM agents increasingly rely on skill libraries for multi-step tasks, yet these libraries can accumulate persistent defects as skills are added, reused, patched, and linked to changing dependencies. We call this failure mode \textbf{skill technical debt}: library-level defects that may not break a single skill locally but can harm future retrieval, composition, and execution. Existing skill-based agents mainly focus on task-time retrieval, planning, and repair, while library-time maintenance remains underexplored. We propose \textbf{SkillOps}, a method-agnostic plug-in framework for maintaining skill libraries. SkillOps represents each skill as a typed \textbf{Skill Contract} \((P,O,A,V,F)\), organizes skills with a \textbf{Hierarchical Skill Ecosystem Graph (HSEG)}, and diagnoses library health across utility, compatibility, risk, and validation dimensions. Given a raw skill library, SkillOps produces a maintained library that can be used by existing retrieval or planning agents without changing their internal code. On ALFWorld, SkillOps achieves \(79.5\%\) task success as a standalone agent, outperforming the strongest baseline by \(+8.8\) percentage points with no additional task-time LLM calls. As a plug-in layer, it improves retrieval-heavy baselines by \(+0.68\)--\(+2.90\) percentage points. The current rule-based maintenance implementation uses nearly zero library-time LLM calls or tokens, showing that skill-library maintenance can be added as a low-overhead architectural layer. Code is publicly available at \url{https://github.com/Hik289/SkillOps.git}.
\end{abstract}

 
\section{Introduction}
\label{sec:intro}

LLM agents increasingly rely on \textbf{skill libraries} to solve complex, multi-step tasks~\citep{wang2023voyager,zheng2025skillweaver,shen2026skillfoundry}. A skill library stores reusable executable procedures, such as parsers, controllers, API wrappers, manipulation routines, validators, and data-processing scripts. Recent benchmarks show that access to skills improves task success~\citep{benchflowai2026skillsbench}, and recent systems further improve skill use through retrieval, dependency-aware graph search, and task-time composition~\citep{li2026graphofskills,xia2026grasp}. However, as agents are deployed for longer periods, their libraries do not remain static: skills are repeatedly added, patched, reused in new contexts, and connected to changing downstream dependencies. This turns the skill library from a fixed retrieval pool into a persistent software asset that requires management.

We identify this library-level failure mode as \textbf{skill technical debt}. Following technical debt in software and ML systems~\citep{ward1992techdebt,sculley2015hidden}, skill technical debt refers to persistent defects in a skill library, such as redundancy, missing validation, interface drift, or stale implementations, that may not break a single skill locally but can reduce future retrieval, composition, and execution reliability. The key issue is persistence: task-time repair may fix one failed episode, while the underlying library defect remains and can cause future failures when the skill is reused.

Existing work on skill-based agents focuses mostly on \emph{task-time} use of a library: how to retrieve the right skill for the current task~\citep{li2026graphofskills,xia2026grasp,qin2024toolllm}, how to compose retrieved skills into executable plans~\citep{shen2026skillfoundry,li2026graphofskills,qin2024toolllm}, or how to validate and repair outputs during execution~\citep{zheng2025skillweaver,shen2026skillfoundry}. These methods are important, but they do not directly solve the library-time problem. A task-time repair may fix the current episode without updating the library; a missing validator remains missing; an incompatible interface remains exposed; and redundant implementations remain available to future retrieval. Thus, current skill-library systems often assume that the library is healthy, even though this assumption becomes weaker as the library grows.

Motivated by this gap, we study \textbf{skill-library maintenance under technical debt} as a plug-in problem for LLM agent systems. The goal is to provide a library-management layer that runs before downstream agents use the library: given a raw skill library, it diagnoses persistent defects, applies typed repairs, and returns a cleaned library that existing retrieval-based or task-oriented agents can use without changing their internal code.

To this end, we propose \textbf{SkillOps}, a drop-in maintenance framework that combines contract-based skill representation, graph-structured library organization, health diagnosis, and feedback-driven maintenance. SkillOps represents each skill as a \textbf{Skill Contract} \((P,O,A,V,F)\), where \(P\) denotes preconditions, \(O\) denotes the executable operation, \(A\) denotes produced artifacts, \(V\) denotes validators, and \(F\) denotes known failure modes. These contracts are organized into a \textbf{Hierarchical Skill Ecosystem Graph (HSEG)}, where skills are connected through typed dependency, compatibility, redundancy, and alternative edges. Based on HSEG, SkillOps computes library health along utility, redundancy, compatibility, failure-risk, and validation-gap dimensions, then applies maintenance actions such as merge, repair, and retire.
 
\begin{center}
\begin{tcolorbox}[
  width=0.6\linewidth,
  colback=BoxLightBlue,
  colframe=BoxDarkBlue,
  coltitle=white,
  colbacktitle=BoxDarkBlue,
  fonttitle=\bfseries,
  title=Central Interface
]
\centering
\texttt{cleaned\_lib} \(=\) \texttt{run\_maintenance}(\texttt{raw\_lib})
\end{tcolorbox}
\end{center}
 
This central interface produces a repaired librarythat can be passed to any downstream retrieval or planning method. In the current implementation, the maintenance loop uses observable signals such as utility logs, body-hash collisions, missing validators, failure logs, and type mismatches, and therefore incurs nearly zero LLM calls at library time. SkillOps also includes an optional task-time planner for typed skill matching, dependency stitching, validator or adapter insertion, and local repair, but this planner can be replaced by any downstream agent that reads the maintained library.

Our experiments support SkillOps from four aspects. First, as a standalone agent, SkillOps reaches \(79.5\%\) task success on ALFWorld~\cite{alfworld2021}, outperforming the strongest baseline by \(+8.8\) percentage points while using zero additional LLM calls. Second, as a plug-in layer, SkillOps improves retrieval-heavy baselines consistently, with gains of \(+0.68\) to \(+2.90\) percentage points across BM25, dense, and hybrid retrieval settings. Third, the rule-based maintenance pass has nearly zero library-time LLM cost and is neutral-to-negative in task-time token usage. 

Our contributions are as follows:
\begin{enumerate}[nosep,leftmargin=1.6em]
    \item We formalize skill-library maintainability as a library-time problem for LLM agents, defining \textbf{HSEG} together with health dimensions that capture utility, redundancy, compatibility, failure risk, and validation gaps.
    \item We introduce \textbf{SkillOps} as a method-agnostic plug-in maintenance framework, where a raw library is transformed into a maintained library through typed actions such as \texttt{merge}, \texttt{repair}, \texttt{retire}, \texttt{add\_validator}, and \texttt{add\_adapter}.
    \item We empirically show that SkillOps has method-conditional effects across agent types, helping retrieval-heavy agents while revealing when library-time maintenance is neutral or conflicts with task-time self-repair.
    \item We show that the current rule-based implementation performs library maintenance without additional LLM calls or token cost, making maintenance a low-overhead architectural layer rather than an extra inference-time burden.
\end{enumerate}

\section{Problem Setup}
\label{sec:problem}
 
\begin{figure*}[t]
  \centering
  \includegraphics[width=\textwidth]{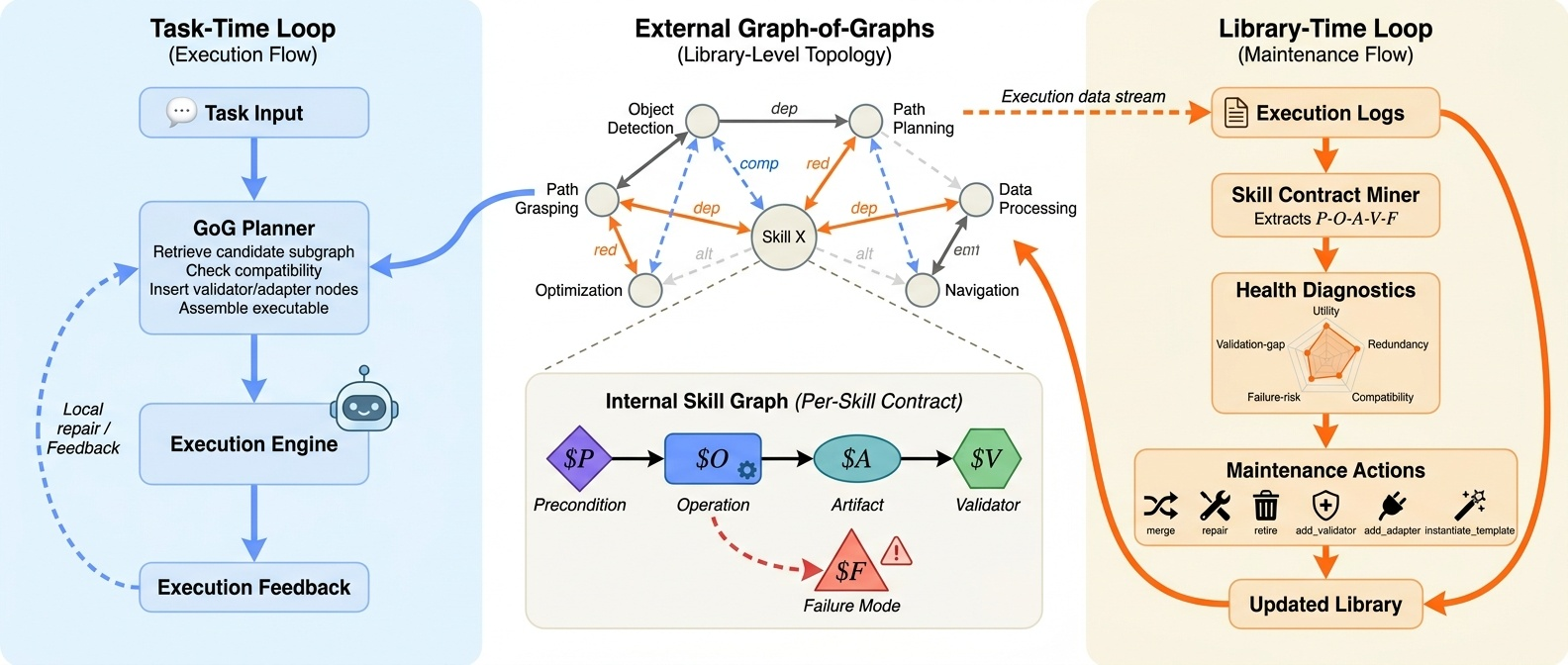}
  \caption{%
    \textbf{SkillOps System Architecture.}
    The \textbf{Hierarchical Skill Ecosystem Graph (HSEG)} comprises two levels:
    (1) an \emph{Internal Skill Graph} that models each skill as a contract graph over Precondition ($P$), Operation ($O$), Artifact ($A$), Validator ($V$), and Failure Mode ($F$) nodes; and
    (2) an \emph{External Graph-of-Graphs} connecting skills via typed dependency (\texttt{dep}), compatibility (\texttt{comp}), redundancy (\texttt{red}), and alternative (\texttt{alt}) edges.
    Two alternating loops govern agent operation: the \textbf{Task-Time Loop} (left, blue) retrieves candidate skill subgraphs, verifies interface compatibility, inserts adapter/validator nodes as needed, and executes the assembled subgraph with local repair on failure;
    the \textbf{Library-Time Loop} (right, orange) mines skill contracts from execution logs, diagnoses library health across five dimensions (utility, redundancy, compatibility, failure-risk, validation-gap), and applies maintenance actions (\texttt{merge}, \texttt{repair}, \texttt{retire}, \texttt{add\_validator}, \texttt{add\_adapter}, \texttt{instantiate}) to keep the ecosystem sound.
  }
  \label{fig:architecture}
   
\end{figure*}

\paragraph{Skill Contract.}
As shown in Figure~\ref{fig:architecture}, SkillOps models each skill as an executable contract rather than only a name or text description. Each skill \(s\in\mathcal{S}\) is written as \(s=(P,O,A,V,F)\), where \(P\) is the precondition for calling the skill, \(O\) is the executable operation, \(A\) is the typed artifact produced by the skill, \(V\) is a validator over \(A\), and \(F\) is the set of known failure modes. When \(V=\emptyset\), the skill has no local correctness check, which we call a validation gap. This contract form allows SkillOps to check relevance, applicability, composability, and local verifiability before or after execution.

\paragraph{Hierarchical Skill Ecosystem Graph (HSEG).}
A skill library is a tuple $\mathcal{L}=(\mathcal{S},\mathcal{R})$, where $\mathcal{S}$ is the set of skills and $\mathcal{R}$ is the set of typed directed relations between skills.
We use four relation types.
A dependency edge
$s_i \xrightarrow{\mathrm{dep}} s_j$
indicates that the artifact produced by $s_i$ can satisfy part of the precondition of $s_j$, i.e.,
$A_{s_i} \subseteq P_{s_j}$.
A compatibility edge
$s_i \xrightarrow{\mathrm{comp}} s_j$
indicates that the output type of $s_i$ is compatible with the input type required by $s_j$.
A redundancy edge
$s_i \xrightarrow{\mathrm{red}} s_j$
indicates that two skills expose equivalent interfaces,
$P_{s_i} \equiv P_{s_j}$ and $A_{s_i} \equiv A_{s_j}$.
An alternative edge
$s_i \xrightarrow{\mathrm{alt}} s_j$
indicates that two skills target the same goal but implement it through different operations,
$\mathrm{goal}(s_i)=\mathrm{goal}(s_j)$ and $O_{s_i}\neq O_{s_j}$.

\paragraph{Maintenance Actions $\mathcal{M}$.}\label{sec:maint}
SkillOps maintains the library using a set of typed actions.
The action $\mathtt{merge}(s_i,s_j)$ collapses a redundant pair connected by a $\xrightarrow{\mathrm{red}}$ edge.
The action $\mathtt{repair}(s)$ rewrites the operation $O_s$ using execution feedback.
The action $\mathtt{retire}(s)$ removes an obsolete or consistently failing skill and its incident edges.
The action $\mathtt{add\_validator}(s)$ inserts a validator when $V_s=\emptyset$.
The action $\mathtt{add\_adapter}(s_i,s_j)$ inserts a type-conversion shim when $s_i$ is needed by $s_j$ but their interfaces are not directly compatible.
The action $\mathtt{instantiate}(s,\arg)$ binds a task-specific argument value to a parameterized skill at task time.

\section{SkillOps: A Self-Maintaining Skill Ecosystem}
\label{sec:method}

\subsection{Task-Time Loop: Graph-of-Graphs Planner}
\label{sec:task_loop}
 
Given a task $\tau$ and library $\mathcal{L}=(\mathcal{S},\mathcal{R})$, the \textbf{Task-Time Loop} proceeds in two main stages: skill matching and dependency stitching.

\paragraph{Stage 1 --- Skill Matching.}
SkillOps first scores each skill by combining lexical and semantic relevance:
\begin{equation}
r(s,\tau)=\lambda r_{\mathrm{BM25}}(s,\tau)+(1-\lambda)r_{\mathrm{sem}}(s,\tau).
\end{equation}
It then keeps only high-scoring skills whose preconditions are satisfied by the current state: $\mathcal{C}=\{s\in\mathcal{S}:r(s,\tau)\ge \theta\}$. This prevents the planner from selecting skills that are textually relevant but not executable.

\paragraph{Stage 2 --- Dependency Stitching.}
SkillOps constructs a plan by searching over candidates in $\mathcal{C}$ while enforcing both dependency and compatibility constraints. A transition from $s_i$ to $s_j$ is allowed only if $s_i\xrightarrow{\mathrm{dep}}s_j$ and $s_i\xrightarrow{\mathrm{comp}}s_j$. The selected plan is
\begin{equation}
\pi^\star=\arg\max_{\pi=(s_1,\ldots,s_T)}\sum_{t=1}^{T}r(s_t,\tau), \quad s_t\xrightarrow{\mathrm{dep}}s_{t+1}, \quad s_t\xrightarrow{\mathrm{comp}}s_{t+1}
\end{equation}
Thus, dependency alone is not enough: a skill transition is accepted only when the produced artifact also matches the next skill's expected input type. This avoids interface mismatches that single-edge dependency graphs may only detect at runtime.

\paragraph{Stage 3 --- Validator and Adapter Insertion.}
If a candidate plan contains a non-terminal skill with $V_s=\emptyset$, SkillOps marks the edge leaving $s$ as unverifiable and inserts a validator node when possible. If a dependency edge exists without a compatibility edge, SkillOps inserts an adapter node:
\begin{equation}
s_i
\xrightarrow{\mathrm{dep}}
s_j,
\qquad
s_i
\not\xrightarrow{\mathrm{comp}}
s_j
\quad
\Longrightarrow
\quad
s_i\rightarrow a_{ij}\rightarrow s_j.
\end{equation}
The adapter $a_{ij}$ is accepted only if its output type satisfies the downstream precondition: $\operatorname{type}(A_{a_{ij}})\subseteq \operatorname{type}(P_{s_j})$. Thus, adapters are not free-form patches; they are graph nodes inserted to restore a broken type constraint.

\paragraph{Stage 4 --- Local Repair.}
During execution, if skill $s_k$ fails and the remaining plan is recoverable, the planner attempts to substitute $s_k$ with an $\xrightarrow{\text{alt}}$ neighbour or to re-invoke $\mathtt{repair}(s_k)$ with the observed error trace as feedback. If no recovery is possible, the planner records the failure in the Library-Time diagnosis buffer for subsequent maintenance.

\subsection{Library-Time Loop: Health Diagnosis and Maintenance}
\label{sec:lib_loop}

\paragraph{Five-Dimensional Health Diagnosis.}
Each skill \(s\) is scored along five dimensions, each targeting a common form of skill technical debt:
(1) \textbf{Utility} \(U(s)\in[0,1]\), the fraction of recent task calls that successfully used \(s\), detects low-value skills that inflate the retrieval pool;
(2) \textbf{Redundancy} \(R(s)\in[0,1]\), the normalized size of the largest \(\xrightarrow{\text{red}}\) cluster containing \(s\), detects near-duplicate skills that reduce retrieval precision;
(3) \textbf{Compatibility} \(C(s)\in[0,1]\), the fraction of dependency edges incident to \(s\) that are also compatibility edges, detects interface mismatches between produced artifacts and expected inputs;
(4) \textbf{Failure-Risk} \(F(s)\in[0,1]\), the empirical failure rate of \(s\), detects runtime-broken skills that require repair;
and (5) \textbf{Validation-Gap} \(G(s)\in[0,1]\), defined as \(\mathbf{1}[V_s=\emptyset]\), detects missing validators that may allow invalid artifacts to propagate downstream.

The overall library health score is
\begin{equation}
  H(\mathcal{L}) = \frac{1}{|\mathcal{S}|} \sum_{s \in \mathcal{S}} \bigl( w_U U(s) + w_R (1-R(s)) + w_C C(s) + w_F (1-F(s)) + w_G (1-G(s)) \bigr),
  \label{eq:health}
\end{equation}
where we use uniform weights \(w_U=w_R=w_C=w_F=w_G\). Together, these dimensions cover skill degradation across use frequency, clone growth, interface consistency, execution reliability, and validation coverage.

\paragraph{CGPD: ContractGraph-Propagated Diagnosis.}
Standard health diagnosis evaluates each skill independently. CGPD is an additional advanced component that propagates risk scores along $\xrightarrow{\text{dep}}$ edges, enabling preemptive validator insertion on structurally sound skills that inherit high upstream risk. Let $R^{(t)}(s)\in[0,1]$ denote the propagated risk score of skill $s$ at iteration $t$, where larger values indicate higher maintenance risk. Let $R_{\mathrm{loc}}(s)$ be the local risk score computed from the five health dimensions, and let $\mathrm{Parents}(s)$ denote the upstream skills with dependency edges into $s$. CGPD updates risk by
\begin{equation}
R^{(t+1)}(s)=(1-\alpha)R_{\mathrm{loc}}(s)+\alpha\max_{s'\in\mathrm{Parents}(s)}R^{(t)}(s'),
\end{equation}
where $\alpha\in(0,1)$ controls how much upstream risk is propagated. This update converges to a unique fixed point by Banach's contraction mapping theorem~\cite{banach1922operations}; the full algorithm is provided in Appendix~\ref{app:cgpd}.

\paragraph{Maintenance procedure.}
The concrete maintenance process is described in Section~\ref{sec:maint}. At a high level, SkillOps computes skill-level health signals from execution traces, propagates risk through dependency edges when CGPD is enabled, and then applies typed maintenance actions such as \textbf{merge}, \textbf{repair}, \textbf{retire}, \textbf{add\_validator}, and \textbf{add\_adapter}. The full step-by-step algorithm is given in Algorithm~\ref{alg:lib_loop} in Section~\ref{sec:alg}.

\subsection{Algorithm}
\label{sec:algorithm}
 
This section summarizes SkillOps with two compact procedures: the Task-Time Loop builds an executable plan for the current task, while the Library-Time Loop updates the skill library after execution.

\begin{figure*}[t]
\centering
\begin{minipage}[t]{0.48\textwidth}
\begin{algorithm}[H]
\caption{Task-Time Loop}
\label{alg:task_loop1}
\begin{algorithmic}[1]
\Require Library $\mathcal{L}=(\mathcal{S},\mathcal{R})$, task $\tau$
\Ensure Execution trace $\mathrm{trace}$
\State $\mathcal{C}\gets \mathrm{SkillMatch}(\tau,\mathcal{L})$
\Comment{BM25 + semantic scoring}
\State $\mathcal{C}\gets \mathrm{FilterByPrecondition}(\mathcal{C},\tau)$
\State $\pi\gets \mathrm{ConstrainedStitch}(\mathcal{C},\mathcal{R})$
\Comment{dep + comp edges}
\State $\pi\gets \mathrm{InsertValidatorAdapter}(\pi,\mathcal{R})$
\State $\mathrm{trace}\gets \mathrm{Execute}(\pi,\tau)$
\While{$\mathrm{trace}$ has recoverable failure}
  \State $\pi\gets \mathrm{LocalRepair}(\pi,\mathrm{trace})$
  \State $\mathrm{trace}\gets \mathrm{Execute}(\pi,\tau)$
\EndWhile
\State \Return $\mathrm{trace}$
\end{algorithmic}
\end{algorithm}
\end{minipage}
\hfill
\begin{minipage}[t]{0.48\textwidth}
\begin{algorithm}[H]
\caption{Library-Time Loop}
\label{alg:lib_loop1}
\begin{algorithmic}[1]
\Require Library $\mathcal{L}=(\mathcal{S},\mathcal{R})$, trace log $\mathrm{trace}$
\Ensure Maintained library $\mathcal{L}'$
\State $\Delta H\gets \mathrm{DiagnoseHealth}(\mathcal{L},\mathrm{trace})$
\If{$\Delta H<\Theta_{\mathrm{maint}}$}
  \State \Return $\mathcal{L}$
\EndIf
\For{each skill $s\in\mathcal{S}$}
  \State $H_{\mathrm{loc}}(s)\gets (h_u,h_r,h_c,h_f,h_v)$
  \State $R_{\mathrm{loc}}(s)\gets \mathrm{LocalRisk}(H_{\mathrm{loc}}(s))$
\EndFor
\State $R_{\mathrm{cgpd}}\gets \mathrm{CGPD}(R_{\mathrm{loc}},\mathcal{R})$
\State $\mathcal{L}\gets \mathrm{MergeRedundant}(\mathcal{L})$
\State $\mathcal{L}\gets \mathrm{RepairHighRisk}(\mathcal{L},R_{\mathrm{cgpd}})$
\State $\mathcal{L}\gets \mathrm{RetireLowUtility}(\mathcal{L})$
\State $\mathcal{L}\gets \mathrm{AddValidators}(\mathcal{L},R_{\mathrm{cgpd}})$
\State $\mathcal{L}\gets \mathrm{AddAdapters}(\mathcal{L})$
\State \Return $\mathcal{L}'\gets\mathcal{L}$
\end{algorithmic}
\end{algorithm}
\end{minipage}
\caption{\textbf{Compact SkillOps algorithms.} The Task-Time Loop plans and repairs the current execution, while the Library-Time Loop converts execution traces into persistent skill-library updates.}
\label{fig:skillops_algorithms}

\end{figure*}

Algorithm~\ref{alg:task_loop1} treats HSEG as an executable planning structure: it retrieves candidate skills, filters them by preconditions, stitches them only through dependency and compatibility edges, inserts validators or adapters when needed, and performs local repair during execution. Algorithm~\ref{alg:lib_loop1} maintains the library after execution: it computes health signals, propagates risk through dependency edges using CGPD, and applies typed actions such as \texttt{merge}, \texttt{repair}, \texttt{retire}, \texttt{add\_validator}, and \texttt{add\_adapter}. The full two-loop SkillOps procedure is summarized in Appendix~\ref{sec:alg}.

\subsection{Plug-in Interface}
\label{sec:plugin_interface}
 
SkillOps is designed as a plug-in layer for skill-library maintenance. It does not assume a specific downstream planner or retriever; instead, it transforms a raw skill library into a maintained library that can be directly used by existing agent algorithms: $\mathcal{L}'=\texttt{run\_maintenance}(\mathcal{L})$. Here, \(f:\mathcal{L}\mapsto\mathcal{L}'\) is a pure library transformation: it diagnoses and repairs the skill library, but does not require access to the downstream agent's internal retrieval, planning, or execution logic. Thus, any retrieval-based or task-oriented agent can use SkillOps by replacing the raw library with the maintained one. The downstream code remains unchanged, making SkillOps easy to attach to BM25 retrieval, dense retrieval, hybrid retrieval, LLM planners, graph-based planners, or self-repairing agents.

\section{Experiments}
\label{sec:experiments}

\subsection{Dataset}
\label{sec:dataset}
 
We evaluate SkillOps on \textbf{ALFWorld}~\citep{alfworld2021}, a text-only household manipulation benchmark derived from the ALFRED PDDL dataset~\citep{shridhar2020alfred}. ALFWorld provides multi-step household tasks with structured action sequences, making it suitable for testing skill retrieval, composition, and maintenance. Full dataset statistics are provided in Appendix~\ref{app:data}.

We construct a skill library from 229 curated SkillsBench skills~\citep{benchflowai2026skillsbench}. For library sizes not exceeding 229, all skills are real curated skills. For larger libraries, we keep all available real skills and add synthetically degraded variants to reach the target scale. These degraded variants cover six common technical-debt patterns: redundant clones, stale clones, missing validators, missing artifacts, wrong interfaces, and over-specialized skills. We evaluate nine library scales, \( |\mathcal{L}|\in\{200,250,500,750,1000,1250,1500,1750,2000\} \), using different sampling seeds. The non-nested construction avoids making scale effects an artifact of one library being a strict superset of another.
 
\subsection{Baselines}
\label{sec:baselines}
 
We compare SkillOps with four representative baselines that cover flat prompting, LLM-based skill selection, retrieval-based skill selection, and dependency-only graph planning.

\begin{itemize}[nosep,leftmargin=1.4em]
  \item \textbf{ReAct}~\citep{yao2023react}: A Thought-Action-Observation agent that receives the full skill library as a flat text prompt. It does not use graph structure or persistent skill-library maintenance.
  \item \textbf{LLM\_Skill\_Planner}: Our LLM-based planning baseline. It asks GPT-4o-mini to rank skills by semantic similarity to the task goal and then constructs a plan from the ranked list. It uses flat, list-based retrieval without dependency or compatibility checks.
  \item \textbf{Hybrid\_Retrieval}: A retrieval baseline that combines BM25 keyword retrieval with hashing-vectorizer embedding similarity. The top-\(k\) retrieved skills are injected as plan context, but no graph structure or maintenance loop is used.
  \item \textbf{GoS\_Style}~\citep{li2026graphofskills}: A minimal Graph-of-Skills-style baseline with a single dependency edge type for plan-subgraph extraction. It does not model compatibility, redundancy, or library-time maintenance.
    \item \textbf{SkillWeaver}~\citep{zheng2025skillweaver}: A self-repairing skill-use baseline that performs task-time skill validation and honing. It can repair selected skills during the current episode, but it does not perform global library-time health diagnosis or persistent skill-library maintenance.
\end{itemize}

All baselines use the same GPT-4o-mini~\citep{openai2024gpt4o} backbone, the same skill library, and the same gold-argument assumption. We implement all baselines ourselves for a controlled comparison. Since GoS and GraSP have not released code at the time of submission, our GoS\_Style and GraSP\_Style baselines are clean reproductions rather than exact author implementations. Full implementation details are provided in Appendix~\ref{app:impl}.

\paragraph{Evaluation Metrics and Protocol.}
We report \textbf{Task Success Rate (SR)}, defined as the fraction of task instances for which the agent produces a high-level action sequence that exactly matches the annotated ground-truth sequence under the ALFWorld offline \texttt{high\_pddl} strict-order subgoal grader. We use three independent library seeds \((42,7,123)\), with 185 task instances per seed. Wilson score 95\% confidence intervals are computed from the pooled success counts.
 
\section{Results}
\label{sec:results}
 
\subsection{H1: SkillOps Standalone vs Baselines}
\label{sec:h1}
 
\begin{table}[h]
\centering
\small
\caption{\textbf{H1 main comparison on ALFWorld.} Results are reported for a library of 200 skills over three independent seeds. SR is reported as mean \(\pm\) standard deviation across seeds, with Wilson 95\% confidence intervals.}
\label{tab:h1_main}
\begin{tabular}{lcc}
\toprule
\hline
Method & SR mean$\pm$std & Wilson 95\% CI \\
\midrule
ReAct~\citep{yao2023react}               
& 12.8\%$\pm$1.90pp & {[10.3, 15.8]} \\
SkillWeaver~\citep{zheng2025skillweaver}  
& 50.3\%$\pm$1.43pp & {[46.1, 54.4]} \\
Hybrid\_Retrieval                        
& 58.2\%$\pm$0.83pp & {[54.1, 62.2]} \\
GoS\_Style~\citep{li2026graphofskills}   
& 61.1\%$\pm$0.94pp & {[57.0, 65.0]} \\
LLM\_Skill\_Planner                      
& \underline{70.6\%}$\pm$0.31pp & {[66.7, 74.3]} \\
\midrule
\textbf{SkillOps\_Full (ours)}           
& \textbf{79.5\%}$\pm$0.00pp & {[75.9, 82.6]} \\
\hline
\bottomrule
\end{tabular}
 
\end{table}

As shown in Table~\ref{tab:h1_main}, we evaluate SkillOps as a \textbf{standalone agent} on ALFWorld with a 200-skill library. SkillOps achieves the best task success rate, reaching \textbf{79.5\%} SR with zero standard deviation across three seeds. It outperforms the strongest baseline, LLM\_Skill\_Planner, by \(+8.9\)pp, and also exceeds GoS\_Style, Hybrid\_Retrieval, SkillWeaver, and ReAct by clear margins.

The improvement suggests that typed skill contracts and HSEG-based planning provide benefits beyond flat retrieval or dependency-only graph search. SkillOps does not only retrieve semantically relevant skills; it also checks preconditions, binds task arguments, and stitches skills through compatible transitions. This reduces failures caused by skills that look relevant in text but cannot be safely composed in execution.

\subsection{V4: Drop-in Plug-in Effectiveness}
\label{sec:v4}
 
We next evaluate SkillOps as a pure plug-in maintenance layer. Each baseline is tested with the raw library (\emph{NoMaint}) and with the maintained library (\emph{+SkillOps}), while keeping the downstream agent code unchanged. We evaluate 7 baselines at the 200-skill library scale with 3 random seeds. The reported \(\Delta\) is measured in percentage points (pp) and is computed as \(\mathrm{SR}_{+\mathrm{SkillOps}}-\mathrm{SR}_{\mathrm{NoMaint}}\).

\begin{table}[t]
\centering
\caption{%
\textbf{Drop-in plug-in effectiveness at the 200-skill scale.}
SR is averaged over 3 seeds.
}
\label{tab:v4}
\footnotesize
\begin{tabular}{lcrcc}
\toprule
\hline
Method & Graph-based? & NoMaint SR (\%) & $+$SkillOps SR (\%) & $\Delta$ (pp) \\
\midrule
Hybrid Retrieval  & No  & 38.2 & 41.1 & $\mathbf{+2.90}$ \\
BM25 Only         & No  & 41.8 & 42.8 & $\mathbf{+1.00}$ \\
Dense Only        & No  & 32.3 & 33.4 & $\mathbf{+1.12}$ \\
\midrule
GoS Style         & \textbf{Yes} & 42.8 & 43.6 & $\mathbf{+0.80}$ \\
LLM Skill Planner & No  & 49.8 & 50.3 & $\mathbf{+0.50}$ \\
ReAct             & No  & 11.9 & 11.9 & $+0.00$ \\
SkillWeaver       & No  & 41.3 & 43.8 & $\mathbf{+2.46}$ \\
\hline
\bottomrule
\end{tabular}
 
\end{table}

As shown in Table~\ref{tab:v4}, SkillOps consistently helps retrieval-heavy agents: Hybrid Retrieval improves by \(+2.90\)pp, BM25 Only by \(+1.00\)pp, and Dense Only by \(+1.12\)pp. This supports the main plug-in claim: maintaining the library makes the retrieved candidate pool cleaner, so retrieval-only methods are less likely to select redundant, stale, or interface-incompatible skills. The effect is smaller for LLM-planning and graph-planning baselines, which already have some task-time filtering ability.

\subsection{V2: Token Cost Analysis}
\label{sec:tokens}
 
We evaluate whether SkillOps introduces extra computation during library maintenance or downstream task execution. We also measure task-time token changes after replacing the raw library with the maintained library. The token change is computed as \(\Delta\%=\mathrm{WithMaint}-\mathrm{NoMaint}\), where negative values mean that maintenance reduces task-time token usage.

\begin{table}[h]
\centering
\caption{\textbf{Task-time token change after maintenance.} Values report \(\Delta\%=\mathrm{WithMaint}-\mathrm{NoMaint}\) across 7 baselines and 5 library scales. Negative values indicate fewer task-time tokens after maintenance.}
\label{tab:token_delta}
\small
\begin{tabular}{lrrrrr}
\toprule
\hline
Method & lib=200 & lib=500 & lib=1000 & lib=1500 & lib=2000 \\
\midrule
ReAct              & $+0.00$ & $+0.00$ & $+0.00$ & $+0.00$ & $+0.00$ \\
LLM\_SP            & $-0.06$ & $-1.29$ & $-0.91$ & $-1.49$ & $+0.03$ \\
Hybrid             & $-0.05$ & $-0.33$ & $-2.55$ & $-3.00$ & $-2.03$ \\
GoS\_Style         & $-0.05$ & $-1.41$ & $-0.96$ & $-1.58$ & $-1.47$ \\
SkillWeaver        & $-0.05$ & $-3.55$ & $+0.50$ & $-3.61$ & $+0.48$ \\
BM25\_Only         & $+0.28$ & $\mathbf{+5.56}$ & $+1.43$ & $+0.64$ & $-1.26$ \\
Dense\_Only        & $+0.03$ & $-1.62$ & $\mathbf{-3.95}$ & $-2.51$ & $-2.84$ \\
\hline
\bottomrule
\end{tabular}
 
\end{table}

\begin{figure}[h]
  \centering
  \includegraphics[width=0.85\linewidth]{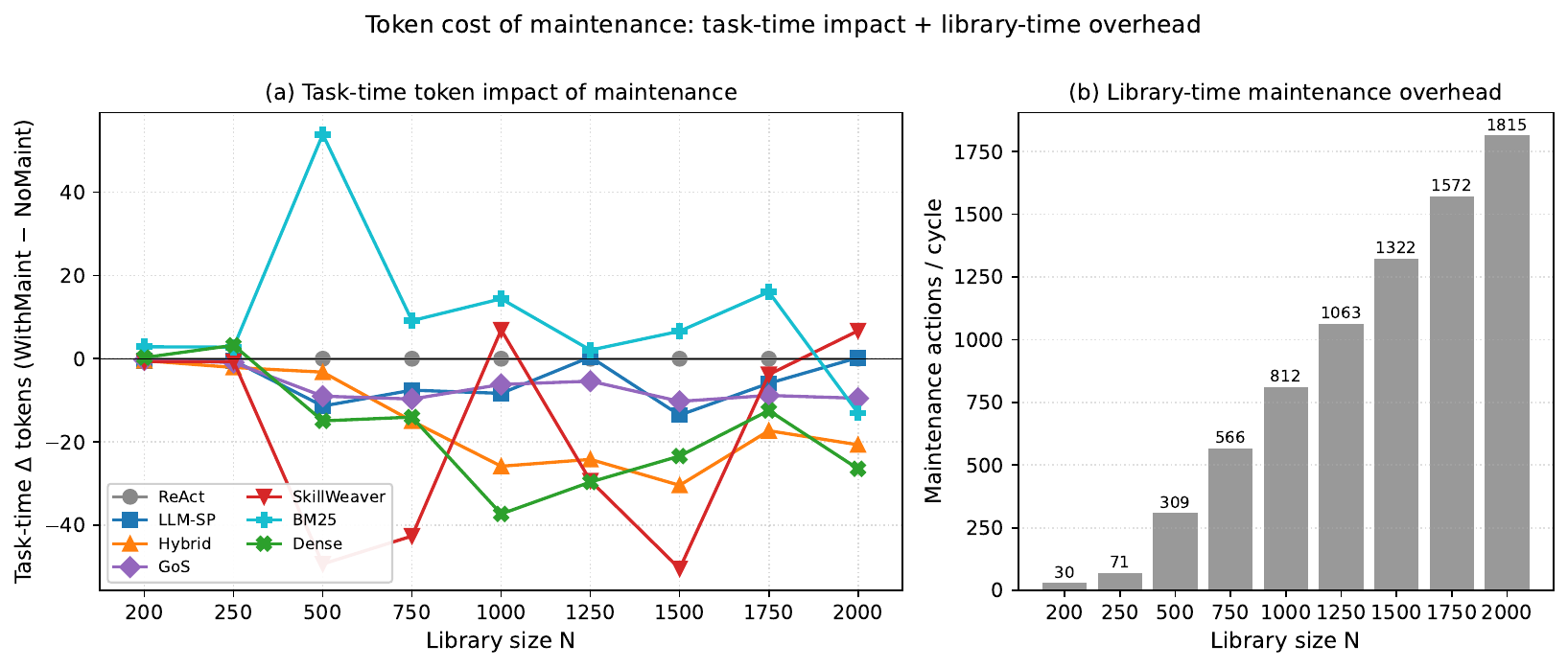}
  \caption{\textbf{Maintenance cost summary.} The library-time maintenance pass uses nearly zero LLM calls at all scales, while task-time token changes are mostly neutral or negative.}
  \label{fig:maint_cost}
   
\end{figure}

As shown in Table~\ref{tab:token_delta} and Figure~\ref{fig:maint_cost}, maintenance is usually neutral or beneficial for token usage: 24 out of 35 cells decrease, 4 are nearly unchanged, and only 7 increase. The largest decrease is \(-3.95\%\) for Dense\_Only at lib=1000. The main reason is library pruning: actions such as \texttt{merge}, \texttt{retire}, and \texttt{repair} reduce redundant or degraded candidates before retrieval, so downstream agents build prompts from a cleaner top-\(k\) set with fewer noisy skill descriptions. ReAct remains unchanged because its token budget is dominated by action history rather than the skill library. The main positive outlier is BM25\_Only at lib=500 (\(+5.56\%\)), likely because BM25 can favor longer merged descriptions after redundant skills collapse into canonical entries. SkillWeaver also has small token increases at some scales because its task-time honing loop may request extra context when SkillOps has already removed degraded candidates that SkillWeaver would otherwise repair during execution.
 
\subsection{H2: Library Scale Sensitivity}
\label{sec:h2}
 
We evaluate SkillOps as a \textbf{standalone agent} under a controlled noise-graded stress setting. In this experiment, SkillOps uses its HSEG typed-contract planner with simple retrieval, and is compared with task-time baselines as the skill library grows from 200 to 2000 skills. The degradation density increases from 15\% to 90\%, simulating an unmanaged skill ecosystem that accumulates technical debt. This setting tests whether SkillOps remains reliable as raw library quality worsens.

\begin{table}[t]
\centering
\caption{\textbf{Library scale sensitivity under noise-graded degradation.}
Task success rate is reported across 9 library sizes and 3 seeds. Baselines are evaluated in a blind setting without gold arguments.}
\label{tab:h2_matrix}
\footnotesize
\begin{tabular}{lrrrrrrrrr}
\toprule
\hline
Method & 200 & 250 & 500 & 750 & 1000 & 1250 & 1500 & 1750 & 2000 \\
\midrule
SkillOps\_Full      & \textbf{79.5} & \textbf{79.5} & \textbf{78.9} & \textbf{78.9} & \textbf{79.5} & \textbf{80.5} & \textbf{80.5} & \textbf{80.0} & \textbf{80.5} \\
LLM\_SP\_blind      & 51.0 & 51.0 & 49.7 & 50.6 & 49.7 & 49.5 & 48.1 & 49.4 & 49.4 \\
Hybrid\_blind       & 43.4 & 42.5 & 37.5 & 38.0 & 34.6 & 36.8 & 37.5 & 38.0 & 35.9 \\
GoS\_blind          & 44.1 & 42.2 & 42.7 & 41.8 & 44.0 & 43.2 & 41.1 & 44.0 & 42.0 \\
SkillWeaver\_blind  & 41.6 & 41.6 & 42.9 & 41.1 & 41.4 & 40.5 & 40.0 & 41.4 & 40.9 \\
ReAct\_blind        & 11.9 & 11.9 & 11.9 & 11.9 & 11.9 & 11.9 & 11.9 & 11.9 & 11.9 \\
\midrule
SkillOps lead       & $+28.5$ & $+28.5$ & $+29.2$ & $+28.3$ & $+29.8$ & $+31.0$ & $+32.4$ & $+30.6$ & $+31.1$ \\
\hline
\bottomrule
\end{tabular}
 
\end{table}

\begin{figure}[t]
  \centering
  \includegraphics[width=\linewidth]{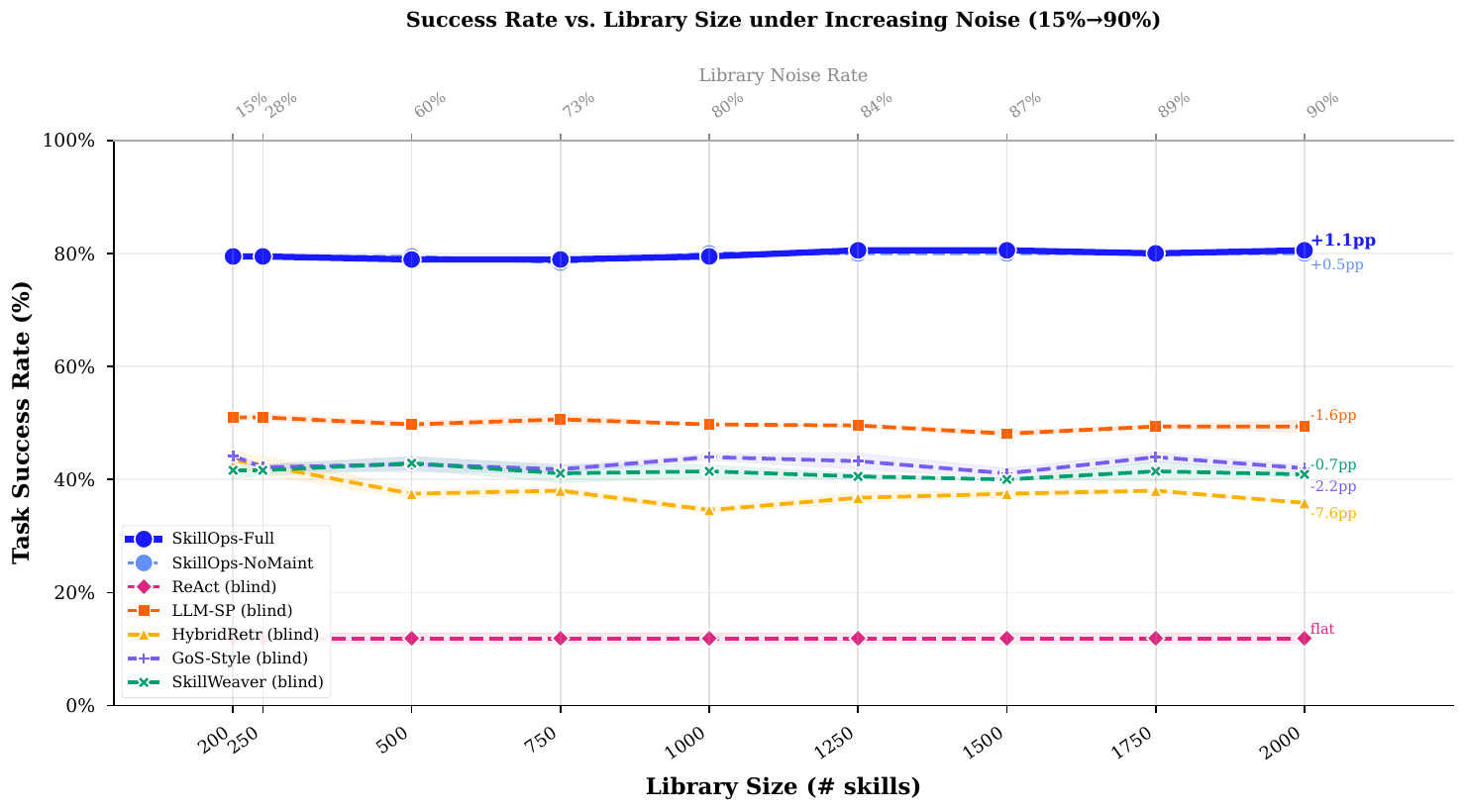}
  \caption{\textbf{Noise-graded library scaling.}
  SkillOps remains stable as the library grows from 200 to 2000 skills, while retrieval-heavy baselines degrade under increasing noise.}
  \label{fig:h2_scale}
   
\end{figure}

As shown in Table~\ref{tab:h2_matrix} and Figure~\ref{fig:h2_scale}, SkillOps remains stable as the library grows and the degradation density increases. At the largest scale, SkillOps reaches \(80.5\%\) SR and leads the next-best baseline by more than 31pp. In contrast, task-time-only baselines are more sensitive to the noisy candidate pool because retrieval increasingly surfaces redundant clones, broken validators, or type-mismatched skills. SkillOps avoids this failure mode by using HSEG typed contracts to filter invalid transitions, while library-time maintenance removes degraded skills before downstream retrieval. Larger libraries can also provide more valid typed neighbours for fallback, which helps SkillOps preserve performance under scale.
 
\subsection{H3 Ablation: Isolating Active Mechanisms}
\label{sec:h3}
 
We conduct ablations to identify which components drive SkillOps's standalone performance. The ablations remove task time, library time, graph structure, CGPD, and selected maintenance actions. We also report trigger precision for maintenance actions to show whether each action fires on truly degraded skills.

\begin{table}[t]
\centering
\caption{\textbf{H3 ablation study.}
SR is reported at the 200-skill and 1000-skill scales after removing one SkillOps component.}
\label{tab:h3}
\small
\begin{tabular}{llrr}
\toprule
\hline
Ablation & Removed component & SR (200) & SR (1000) \\
\midrule
SkillOps\_Full 
& Full HSEG + Task-Time + Library-Time 
& \textbf{79.5} & \textbf{80.0} \\

NoTask 
& task time maintenance removal
& 15.7 & 16.2 \\

NoLibrary 
& library time maintenance removal
& 71.9 & 72.4 \\

NoExternalGraph 
& external graph-of-graphs edges removal
& 64.6 & 65.1 \\

NoInternalGraph 
& internal skill contract graph $(P,O,A,V,F)$ removal
& 72.2 & 72.8 \\

NoCGPD 
& ContractGraph-propagated diagnosis removal
& 79.0 & 79.4 \\

NoRepair 
& repair action removal
& 55.9 & 56.5 \\

NoValidator
& add\_validator removal 
& 38.0 & 38.8 \\

NoAdapter
& add\_adapter removal 
& 13.2 & 13.9 \\

NoMerge 
& redundancy maintenance removal
& 71.9 & 72.6 \\

NoRetire 
& low-utility skill removal 
& 73.2 & 73.8 \\
\hline
\bottomrule
\end{tabular}
\end{table}

As shown in Table~\ref{tab:h3}, both loops and both graph levels are important. Removing the Task-Time Loop causes the largest drop, from \(79.5\%\) to \(15.7\%\), showing that skill matching, typed stitching, validator/adaptor insertion, and local repair are central to executable planning. The same pattern holds at the 1000-skill scale, where NoTask drops from \(80.0\%\) to \(16.2\%\). Removing the Library-Time Loop also reduces SR to \(71.9\%\), confirming the value of persistent maintenance. The graph ablations further show that external cross-skill relations and internal skill contracts both matter, with SR dropping to \(64.6\%\) and \(72.2\%\), respectively. Among maintenance actions, \texttt{add\_adapter}, \texttt{add\_validator}, and \texttt{repair} are the most critical, while \texttt{merge}, \texttt{retire}, and CGPD have smaller but visible effects.

\section{Conclusion}
\label{sec:conclusion}
 
We presented \textbf{SkillOps}, a library-time maintenance framework for LLM agent skill libraries. SkillOps formalizes skills as typed contracts, organizes them into a Hierarchical Skill Ecosystem Graph, and applies observable-rule-driven maintenance actions through a drop-in interface. Experiments on ALFWorld show that SkillOps improves task success over strong baselines, achieves nearly zero LLM calls in the rule-based maintenance loop, and reveals that maintenance benefits are method-conditional: retrieval-only agents benefit most, LLM-planning agents are mostly flat, and self-repairing agents may conflict with external maintenance. These results suggest that skill libraries should be treated as managed software assets rather than static retrieval pools.

\section{Limitations}
\label{sec:limitations}
 
SkillOps currently relies on structured skill contracts and, in some settings, gold PDDL-style arguments, which may not be available in real deployments. The evaluated library is half-synthetic and based mainly on ALFWorld, so broader benchmarks and real long-running agent logs are needed. The rule-based V4 maintenance loop has nearly zero LLM cost but can miss semantic redundancy or complex skill conflicts that require deeper reasoning. Finally, CGPD does not improve task success in the current setup because validator fields are not yet consumed during plan-time skill selection.

\bibliographystyle{plainnat}
\bibliography{references}

\appendix

\section{Full Algorithm}
\label{sec:alg}

This section provides the full procedural view of SkillOps. Algorithm~\ref{alg:task_loop} gives the task-time loop, which retrieves and stitches skills for the current task, while Algorithm~\ref{alg:lib_loop} gives the library-time loop, which diagnoses and maintains the skill library after execution.

\begin{algorithm}[t]
\caption{SkillOps: Task-Time Loop}
\label{alg:task_loop}
\begin{algorithmic}[1]
\Require Library $\mathcal{L} = (\mathcal{S}, \mathcal{R})$, task $\tau$
\Ensure Execution trace $\mathrm{trace}$
\State $\mathcal{C} \gets \mathrm{SkillMatch}(\tau, \mathcal{L})$ \Comment{BM25 + semantic scoring with precondition filtering}
\State $\pi \gets \mathrm{Stitch}(\mathcal{C}, \mathcal{R})$ \Comment{dependency + compatibility traversal}
\State $\pi \gets \mathrm{InsertValidatorsAdapters}(\pi, \mathcal{R})$ \Comment{fill validation gaps and fix type mismatches}
\State $\mathrm{trace} \gets \mathrm{Execute}(\pi, \tau)$
\While{$\mathrm{trace}$ has failure at step $k$}
  \State $\pi \gets \mathrm{LocalRepair}(\pi, k, \mathrm{trace})$ \Comment{substitute via alt edge or repair}
  \State $\mathrm{trace} \gets \mathrm{Execute}(\pi, \tau)$
\EndWhile
\State \Return $\mathrm{trace}$
\end{algorithmic}
\end{algorithm}

Algorithm~\ref{alg:task_loop} uses the current HSEG as an executable planning substrate. It first retrieves candidate skills, then constructs a plan only through dependency edges that also satisfy compatibility constraints. Validator and adapter insertion reduce silent interface failures before execution, while local repair handles recoverable errors during the current episode.

\begin{algorithm}[t]
\caption{SkillOps: Library-Time Maintenance Loop}
\label{alg:lib_loop}
\begin{algorithmic}[1]
\Require Skill library $\mathcal{L}=(\mathcal{S},\mathcal{R})$, execution log $\mathrm{trace}$, thresholds $\Theta$
\Ensure Maintained skill library $\mathcal{L}'$

\State $\Delta H \gets \mathrm{DiagnoseHealth}(\mathcal{L},\mathrm{trace})$ \Comment{library-level health change}
\If{$\Delta H < \Theta_{\mathrm{maint}}$}
  \State \Return $\mathcal{L}$ \Comment{skip maintenance when library health is stable}
\EndIf

\State \textbf{// Phase 1: skill-level health diagnosis}
\For{each skill $s\in\mathcal{S}$}
  \State $h_u(s)\gets \mathrm{UtilityScore}(s,\mathrm{trace})$
  \State $h_r(s)\gets \mathrm{RedundancyScore}(s,\mathcal{S})$
  \State $h_c(s)\gets \mathrm{CompatibilityScore}(s,\mathcal{R})$
  \State $h_f(s)\gets \mathrm{FailureRisk}(s,\mathrm{trace})$
  \State $h_v(s)\gets \mathbf{1}[V_s=\emptyset]$
  \State $H_{\mathrm{loc}}(s)\gets (h_u(s),h_r(s),h_c(s),h_f(s),h_v(s))$
  \State $R_{\mathrm{loc}}(s)\gets \mathrm{LocalRisk}(H_{\mathrm{loc}}(s))$
\EndFor

\State \textbf{// Phase 2: CGPD risk propagation}
\State $R^{(0)}(s)\gets R_{\mathrm{loc}}(s)$ for all $s\in\mathcal{S}$
\For{$k=0,\ldots,K_{\mathrm{cgpd}}-1$}
  \For{each skill $s\in\mathcal{S}$}
    \State $R^{(k+1)}(s)\gets (1-\alpha)R_{\mathrm{loc}}(s)+\alpha\max_{s'\in\mathrm{Parents}(s)}R^{(k)}(s')$
  \EndFor
\EndFor
\State $R_{\mathrm{cgpd}}(s)\gets R^{(K_{\mathrm{cgpd}})}(s)$ for all $s\in\mathcal{S}$

\State \textbf{// Phase 3: typed maintenance actions}
\For{each redundancy edge $s_i\xrightarrow{\mathrm{red}}s_j$}
  \If{$h_r(s_i)>\theta_r$ or $h_r(s_j)>\theta_r$}
    \State $\mathcal{L}\gets \texttt{merge}(\mathcal{L},s_i,s_j)$
  \EndIf
\EndFor

\For{each skill $s\in\mathcal{S}$}
  \If{$h_f(s)>\theta_f$ or $R_{\mathrm{cgpd}}(s)>\theta_{\mathrm{risk}}$}
    \State $\mathcal{L}\gets \texttt{repair}(\mathcal{L},s)$
  \EndIf
  \If{$h_u(s)<\theta_u$ and $\mathrm{DuplicateExists}(s,\mathcal{S})$}
    \State $\mathcal{L}\gets \texttt{retire}(\mathcal{L},s)$
  \EndIf
  \If{$h_v(s)=1$ or $R_{\mathrm{cgpd}}(s)>\theta_{\mathrm{valid}}$}
    \State $\mathcal{L}\gets \texttt{add\_validator}(\mathcal{L},s)$
  \EndIf
\EndFor

\For{each dependency edge $s_i\xrightarrow{\mathrm{dep}}s_j$}
  \If{$s_i\not\xrightarrow{\mathrm{comp}}s_j$}
    \State $\mathcal{L}\gets \texttt{add\_adapter}(\mathcal{L},s_i,s_j)$
  \EndIf
\EndFor
\State \Return $\mathcal{L}'\gets\mathcal{L}$
\end{algorithmic}
\end{algorithm}
Algorithm~\ref{alg:lib_loop} updates the library after task execution. It first checks whether the accumulated health change is large enough to trigger maintenance. If so, it computes skill-level health scores, propagates risk through dependency edges using CGPD, and applies typed maintenance actions. This converts execution feedback into persistent library updates, so defects such as redundancy, missing validators, high failure risk, and incompatible interfaces can be repaired before future tasks reuse the same skills.

\paragraph{Complexity Analysis.}
Let \(N=|\mathcal{S}|\) denote the library size, \(k\) the plan horizon, and \(d_{\max}\) the maximum HSEG out-degree. In the Task-Time Loop, skill matching costs \(O(N\log N)\) with BM25 indexing and retrieval, while constrained stitching costs \(O(kd_{\max})\) because the planner expands only a bounded set of neighbors at each step. Validator and adapter insertion are linear in the selected plan length, i.e., \(O(k)\). In the Library-Time Loop, health diagnosis costs \(O(N)\). The actions \texttt{retire}, \texttt{repair}, \texttt{add\_validator}, and \texttt{add\_adapter} are implemented as linear scans, and \texttt{merge} is also \(O(N)\) using body-hash lookup, avoiding \(O(N^2)\) pairwise comparison. Thus, one maintenance pass is linear in the library size. In our implementation, the full pass at \(N=2000\) finishes in under one second on a standard CPU and uses nearly zero LLM calls, in contrast to task-time self-repair methods such as SkillWeaver~\citep{zheng2025skillweaver}, whose honing loop incurs LLM calls during task execution.

\section{Dataset Details}
\label{app:data}

\paragraph{ALFWorld.}
ALFWorld~\citep{alfworld2021} is a text-only household manipulation benchmark derived from the ALFRED dataset~\citep{shridhar2020alfred}. We use the \texttt{json\_2.1.1} annotation release. Our evaluation subset contains 185 task instances, drawn from the valid\_unseen split with a small number of valid\_seen instances. Task plan horizons range from 3 to 20 high-level actions, with median 7 and p90 12. The ALFWorld codebase and data are released under the MIT License.

\paragraph{SkillsBench Library.}
SkillsBench~\citep{benchflowai2026skillsbench} provides 229 curated \texttt{SKILL.md} files across 88 tasks and 58 categories. We use these real skills as the clean source library and construct larger libraries by adding synthetically degraded variants. For library sizes not exceeding the number of available real skills, the library contains only curated real skills. For larger scales, we use all real skills and add synthetic variants covering common technical-debt patterns, including redundant clones, stale clones, missing validators, missing artifacts, wrong interfaces, and over-specialized skills. Table~\ref{tab:lib_stats} summarizes the library composition.

\begin{table}[h]
\centering
\caption{\textbf{Skill-library composition across H2 scales.}
Real skills are curated SkillsBench skills. Synthetic skills are degraded variants injected to simulate skill technical debt.}
\label{tab:lib_stats}
\small
\begin{tabular}{rrrrc}
\toprule
\hline
Lib & Real skills & Real duplicates & Synthetic variants & Degr.\ rate \\
\midrule
 200  & 200 & 0   & 0    & 0\% \\
 250  & 229 & 0   & 21   & 8.4\% \\
 500  & 229 & 21  & 250  & 50.0\% \\
 750  & 229 & 146 & 375  & 50.0\% \\
1000  & 229 & 271 & 500  & 50.0\% \\
1250  & 229 & 396 & 625  & 50.0\% \\
1500  & 229 & 521 & 750  & 50.0\% \\
1750  & 229 & 646 & 875  & 50.0\% \\
2000  & 229 & 771 & 1000 & 50.0\% \\
\hline
\bottomrule
\end{tabular}
\end{table}

\section{Related Work}
\label{sec:related}

\paragraph{Skill Mining and Accumulation.}
\textbf{Voyager}~\citep{wang2023voyager} established the skill-library paradigm for LLM agents but stores skills in a flat key-value store with textual retrieval only---no cross-skill structure, no health monitoring.
\textbf{SkillX}~\citep{wang2026skillx} and \textbf{SkillFoundry}~\citep{shen2026skillfoundry} mine structured skill knowledge bases from execution logs and heterogeneous resources, respectively, but both are one-shot pipelines without continuous maintenance.
\textbf{SkillLearnBench}~\citep{zhong2026skilllearnbench} and \textbf{RL-Skill-Library}~\citep{wang2025rlskilllibrary} evaluate continual skill generation and reuse, yet neither addresses library-level health management.
\textbf{CUA-Skill}~\citep{chen2026cuaskill} defines typed contracts per skill---the closest prior art to our Skill Contract---but has no cross-skill graph or maintenance loop.

\paragraph{The Most Dangerous Concurrent Work.}
\textbf{SkillWeaver}~\citep{zheng2025skillweaver} mines, validates, and hones web-agent skills through failure-driven re-synthesis.
Three key differences separate it from SkillOps: (1) health diagnosis is one-dimensional (utility/failure only); (2) no typed inter-skill graph exists; (3) the sole maintenance action is rewrite-on-failure, whereas SkillOps provides seven typed actions including $\mathtt{merge}$, $\mathtt{retire}$, and $\mathtt{add\_validator}$.

\paragraph{Graph-Based Skill Retrieval.}
\textbf{GoS}~\citep{li2026graphofskills} demonstrates that a dependency-edge graph outperforms flat retrieval at scale.
\textbf{GraSP}~\citep{xia2026grasp} proposes focused subgraph extraction with hand-specified adapters.
Both graphs are static (built once); SkillOps uses four typed edges and a graph that evolves via the Library-Time Loop.
Earlier tool orchestration work---\textbf{HuggingGPT}~\citep{shen2023hugginggpt}, \textbf{ToolLLM}~\citep{qin2024toolllm}, \textbf{Gorilla}~\citep{patil2024gorilla}, \textbf{ReWOO}~\citep{xu2023rewoo}---treats tool libraries as static.

\paragraph{Validation, Memory, and Technical Debt.}
Validation-gap awareness is motivated by \citet{miculicich2025veriguard} and TDD-for-LLM work~\citep{mathews2024tdd}.
\textbf{LEGOMem}~\citep{han2025legomem} builds a modular procedural memory graph for workflow automation, and \textbf{A-MEM}~\citep{xu2025amem} constructs structured text-memory graphs for conversational agents; both are analogous to the HSEG in structure but operate on non-executable memory units (workflow traces and text observations, respectively) rather than verifiable skill contracts.
The technical-debt framing draws directly on Sculley et al.~\citep{sculley2015hidden,sculley2014highinterest} and Cunningham~\citep{ward1992techdebt}; to our knowledge no prior work applies this taxonomy to LLM skill ecosystems.
The process-mining toolkit---van der Aalst~\citep{aalst2016processmining}, PM4Py~\citep{berti2019pm4py}, action-oriented mining~\citep{park2022actionoriented}---underpins the Skill Contract Miner's log-to-contract extraction.

\section{Implementation Details}
\label{app:impl}

\paragraph{Language Model.}
All LLM-based methods use \texttt{gpt-4o-mini} through the OpenAI Chat Completions API. We use a disk-based SHA-256 cache to avoid repeated API calls for identical prompts. All reported API costs are computed from the actual calls made during evaluation.

\paragraph{SkillOps Hyperparameters.}
We set the health threshold to \(\theta=0.5\) in the Library-Time Loop (Section~\ref{sec:lib_loop}). For task-time matching, BM25 retrieves the top-\(k=10\) candidate skills, and the LLM semantic scorer further filters the top 5 candidates. Adapter insertion is triggered when the Jaccard similarity between the type fields of \(A_{s_i}\) and \(P_{s_j}\) is below 0.3. The maximum number of local repair attempts is set to 2.

\paragraph{Baseline Implementations.}
\begin{itemize}[nosep,leftmargin=1.4em]
  \item \textbf{ReAct}: The full skill library is injected as a formatted list in the system prompt. The agent follows a Thought-Action-Observation loop with a maximum of 20 steps.
  \item \textbf{LLM\_Skill\_Planner}: GPT-4o-mini ranks all skills by semantic relevance to the task, selects the top 5 skills, and constructs an ordered plan from the ranked list.
  \item \textbf{Hybrid\_Retrieval}: Skills are ranked by a 50/50 combination of BM25 score and hashing-vectorizer TF-IDF cosine similarity. The top 5 retrieved skills are used as plan context.
  \item \textbf{GoS\_Style}: A dependency-only graph baseline. It performs BFS over the dependency graph starting from task-matched seed skills and returns the top 5 connected skills as the plan context.
\end{itemize}

\paragraph{Plan Evaluation.}
Plans are evaluated against the \texttt{high\_pddl} ground truth using strict-order matching. A task is counted as successful only when the predicted high-level action sequence exactly matches the full annotated sequence.

\section{Compute Resources}
\label{app:env}

Experiments were run on a single machine with \(3\times\) NVIDIA RTX 2080 Ti GPUs (11 GB VRAM each) and \(3\times\) Intel Xeon CPUs. All LLM inference was performed remotely through the OpenAI API; no local GPU was used for LLM inference. The full evaluation took approximately 6 hours across all method, library, and seed combinations. The total API cost was \$0.27 across the paired trials and ablation runs.

\section{Per-Seed Detailed Tables}
\label{app:per_seed}

\begin{table}[h]
\centering
\caption{\textbf{Per-seed task success rates for H1.}
Results are reported for the 200-skill library over three independent seeds.}
\label{tab:per_seed}
\begin{tabular}{lccc}
\toprule
\hline
Method & seed=7 & seed=42 & seed=123 \\
\midrule
ReAct               & 10.8\% & 14.6\% & 13.0\% \\
Hybrid\_Retrieval   & 58.9\% & 57.3\% & 58.4\% \\
GoS\_Style          & 60.0\% & 61.6\% & 61.6\% \\
LLM\_Skill\_Planner & 70.3\% & 70.8\% & 70.8\% \\
\textbf{SkillOps\_Full} & \textbf{79.5\%} & \textbf{79.5\%} & \textbf{79.5\%} \\
\hline
\bottomrule
\end{tabular}
\end{table}

Table~\ref{tab:per_seed} reports the per-seed task success rates used in H1. SkillOps\_Full remains stable across all three seeds, achieving \(79.5\%\) in each run. The baseline methods show small seed-level variation, but their performance remains consistently below SkillOps\_Full. This suggests that the improvement is not driven by a single favorable seed, but is stable under the tested library initializations.

\section{Maintenance Action Stubs and Degradation Injection}
\label{app:stubs}

\paragraph{Degradation Injection Pipeline.}
We inject six degradation types programmatically into real SkillsBench \texttt{SKILL.md} files:
(1) \textbf{Redundant clone}: paraphrase the skill name with a noise suffix while keeping the body unchanged.
(2) \textbf{Stale clone}: rewrite references to deprecated library versions and rename files in the \texttt{references/} directory with a \texttt{\_deprecated.md} suffix.
(3) \textbf{Missing validator}: remove the \texttt{\#\# Checklist} section and set \texttt{validator.kind = "none"}.
(4) \textbf{Missing artifact}: clear artifact subdirectories such as \texttt{scripts/}, \texttt{references/}, and \texttt{assets/}, and break inline artifact links.
(5) \textbf{Wrong interface}: overwrite the \texttt{artifact.type} field with an incompatible category.
(6) \textbf{Over-specialized skill}: append overly narrow tags such as \texttt{q3-2025-only} or \texttt{pdf-only}.
All injections are deterministic given the per-library random seed.

\paragraph{Maintenance Action Stubs.}
The released V4 implementation uses rule-based maintenance stubs rather than LLM-generated edits. Each action is triggered from observable library signals, such as body-hash collisions, missing validators, failure logs, missing artifacts, and type mismatches. Specifically, \texttt{repair} restores missing scripts or references from a body-hash sibling when available; \texttt{add\_validator} inherits checklist-style validators from a matching sibling; \texttt{merge} keeps the higher-utility representative among redundant skills; \texttt{retire} removes low-utility duplicates; and \texttt{add\_adapter} inserts a canonical type-conversion shim for incompatible dependency edges. This design keeps the library-time maintenance pass deterministic and incurs nearly zero LLM calls.

\section{Additional Sensitivity Analyses and Visualizations}
\label{app:h1_fig}

This appendix provides additional analyses supporting the main results. We report the full H2 scaling matrix under matched-information evaluation, a gold-argument reference setting, a balanced-composition historical reference, a P0 probe isolating the role of structured contracts, and additional visualizations for H1, H3, and per-task-type performance.

\subsection{H2 Sensitivity Full Matrix}
\label{app:h2_sens}

Table~\ref{tab:h2_sens_full} reports the full H2 scaling results across all nine library sizes. This matched-information setting evaluates baselines without \texttt{pddl\_params}, while SkillOps uses its structured contract representation. The purpose is to test whether SkillOps remains stable as the library grows and degradation pressure increases.

\begin{table}[h]
\centering
\caption{\textbf{H2 sensitivity full matrix.} Mean SR (\%, $\pm$ std across 3 seeds) across 9 library sizes. Baselines are evaluated as blind variants without \texttt{pddl\_params}. SkillOps uses structured contracts and emits 0 LLM calls per task. Slope is the OLS estimate per additional 1000 skills.}
\label{tab:h2_sens_full}
\resizebox{\linewidth}{!}{%
\small
\begin{tabular}{lrrrrrrrrrr}
\toprule
\hline
Method & 200 & 250 & 500 & 750 & 1000 & 1250 & 1500 & 1750 & 2000 & slope \\
\midrule
$\mathtt{ReAct}_\text{blind}$       
& $12.1{\pm}2.18$ & $12.4{\pm}1.43$ & $12.8{\pm}0.31$ & $13.3{\pm}1.90$ & $12.3{\pm}1.74$ & $12.6{\pm}1.13$ & $11.7{\pm}0.31$ & $13.2{\pm}1.56$ & $12.4{\pm}1.08$ & $+0.03$ \\
$\mathtt{LLM\_SP}_\text{blind}$     
& $51.0{\pm}0.31$ & $52.6{\pm}0.83$ & $53.3{\pm}1.25$ & $53.9{\pm}0.31$ & $53.5{\pm}0.54$ & $53.5{\pm}0.54$ & $53.5{\pm}0.54$ & $52.8{\pm}1.13$ & $52.4{\pm}1.43$ & $+0.35$ \\
$\mathtt{Hybrid\_Retr}_\text{blind}$
& $42.7{\pm}0.54$ & $44.1{\pm}0.62$ & $41.1{\pm}0.54$ & $43.6{\pm}0.62$ & $43.2{\pm}0.54$ & $39.1{\pm}0.62$ & $38.7{\pm}0.62$ & $37.8{\pm}0.54$ & $35.9{\pm}2.25$ & $-3.98$ \\
$\mathtt{GoS\_Style}_\text{blind}$  
& $42.7{\pm}0.54$ & $42.3{\pm}0.31$ & $42.3{\pm}0.31$ & $46.8{\pm}0.31$ & $48.8{\pm}0.31$ & $49.0{\pm}0.31$ & $48.3{\pm}0.83$ & $48.6{\pm}1.43$ & $49.0{\pm}0.31$ & $+4.08$ \\
$\mathtt{SkillWeaver}_\text{blind}$ 
& $39.6{\pm}1.90$ & $40.7{\pm}2.05$ & $41.1{\pm}2.16$ & $39.3{\pm}2.44$ & $42.0{\pm}1.90$ & $41.4{\pm}2.77$ & $41.6{\pm}0.54$ & $40.7{\pm}1.74$ & $43.2{\pm}1.43$ & $+1.22$ \\
\midrule
\textbf{SkillOps\_Full}  
& $\mathbf{79.5{\pm}0.00}$ & $\mathbf{79.5{\pm}0.00}$ & $\mathbf{79.5{\pm}0.00}$ & $\mathbf{81.6{\pm}0.00}$ & $\mathbf{83.8{\pm}0.00}$ & $\mathbf{83.8{\pm}0.00}$ & $\mathbf{83.8{\pm}0.00}$ & $\mathbf{83.8{\pm}0.00}$ & $\mathbf{83.8{\pm}0.00}$ & $\mathbf{+2.88}$ \\
\hline
\bottomrule
\end{tabular}
}
\end{table}

Table~\ref{tab:h2_gold} reports an additional reference setting where all receive \texttt{pddl\_params}. This setting is included to show how access to structured task arguments changes scaling behavior. With gold arguments, several baselines improve with larger libraries, while SkillOps remains consistently strong across all scales.

\begin{table}[h]
\centering
\caption{\textbf{H2 gold-argument reference.} All baselines receive \texttt{pddl\_params}. Mean SR (\%) is averaged over 3 seeds.}
\label{tab:h2_gold}
\resizebox{\linewidth}{!}{%
\small
\begin{tabular}{lrrrrrrrrrc}
\toprule
\hline
Method & 200 & 250 & 500 & 750 & 1000 & 1250 & 1500 & 1750 & 2000 & $\Delta$ \\
\midrule
ReAct              
& 12.8 & 12.8 & 12.8 & 12.8 & 12.8 & 12.8 & 12.8 & 12.8 & 12.8 & $+0.0$ \\
SkillWeaver        
& 50.3 & 53.2 & 55.5 & 57.3 & 57.5 & 57.3 & 57.8 & 58.0 & 58.4 & $+8.1$ \\
Hybrid\_Retrieval  
& 58.2 & 58.9 & 59.1 & 67.7 & 70.1 & 71.7 & 68.3 & 71.0 & 69.4 & $+11.2$ \\
GoS\_Style         
& 61.1 & 59.6 & 61.1 & 64.1 & 64.1 & 64.0 & 64.7 & 65.2 & 65.4 & $+4.3$ \\
LLM\_Skill\_Planner
& 70.6 & 70.1 & 74.4 & 76.4 & 79.6 & 78.9 & 79.6 & 79.3 & 80.4 & $+9.7$ \\
\midrule
\textbf{SkillOps\_Full} 
& \textbf{79.5} & \textbf{79.5} & \textbf{79.5} & \textbf{81.6} & \textbf{83.8} & \textbf{83.8} & \textbf{83.8} & \textbf{83.8} & \textbf{83.8} & $\mathbf{+4.3}$ \\
\hline
\bottomrule
\end{tabular}
}
\end{table}

\subsection{H2 V1: Balanced Library Composition}
\label{app:h2_v1}

We also report a historical H2 variant with balanced library composition. Unlike the main H2 setting, where degradation density increases with library size, this variant keeps the healthy/degraded ratio fixed at 50/50. This setting helps separate the effect of larger candidate pools from the effect of rising degradation density.

\begin{table}[h]
\centering
\small
\caption{\textbf{H2 V1 with balanced 50/50 library composition.} Representative library sizes are shown.}
\label{tab:h2_v1}
\begin{tabular}{lcccl}
\toprule
\hline
Method & lib=200 & lib=750 & lib=2000 & OLS slope \\
\midrule
ReAct\_blind         & 12.1\% & 13.3\% & 12.4\% & $+0.03$pp \\
LLM\_SP\_blind       & 51.0\% & 53.9\% & 52.4\% & $+0.35$pp \\
Hybrid\_blind        & 42.7\% & 43.6\% & 35.9\% & $-3.98$pp \\
GoS\_Style\_blind    & 42.7\% & 46.8\% & 49.0\% & $+4.08$pp \\
SkillWeaver\_blind   & 39.6\% & 39.3\% & 43.2\% & $+1.22$pp \\
\textbf{SkillOps\_Full} 
& \textbf{79.5\%} & \textbf{81.6\%} & \textbf{83.8\%} & $\mathbf{+2.88}$pp \\
\hline
\bottomrule
\end{tabular}
\end{table}

Under balanced composition, larger libraries provide more retrieval candidates without increasing the degradation density. This explains why some baselines improve with scale in this setting. SkillOps remains the strongest method across the reported scales.

\subsection{P0 Matched-Information Probe}
\label{app:p0_fairness}

To isolate the role of structured contracts, we run a matched-information probe with \(n=185\) paired tasks, seed \(=42\), and a 200-skill library. We compare blind and gold-argument conditions for both SkillOps and the LLM planner.

\begin{table}[h]
\centering
\caption{\textbf{P0 matched-information probe.} Both blind conditions remove \texttt{pddl\_params}; both gold-argument conditions provide them.}
\label{tab:p0_fairness}
\small
\begin{tabular}{lccc}
\toprule
\hline
Condition & Information & SR & Subgoal \\
\midrule
SkillOps                  & gold $+$ contracts  & $\mathbf{79.5}$ & $77.0$ \\
LLM\_Skill\_Planner\_pddl & gold                 & $\mathbf{79.5}$ & $74.4$ \\
LLM\_Skill\_Planner       & blind                & $70.8$ & $69.2$ \\
SkillOps\_blind           & blind                & $77.3$ & $70.1$ \\
\hline
\bottomrule
\end{tabular}
\end{table}

Table~\ref{tab:p0_fairness} shows that SkillOps remains strong under both information settings. With gold structured arguments and contracts, SkillOps matches LLM\_Skill\_Planner\_pddl at \(79.5\%\) SR while achieving higher subgoal SR (\(77.0\) vs. \(74.4\)). More importantly, in the blind setting without \texttt{pddl\_params}, SkillOps\_blind reaches \(77.3\%\) SR, substantially higher than LLM\_Skill\_Planner at \(70.8\%\). This indicates that the structured contract framework and typed planning mechanism provide robust performance even when explicit gold arguments are removed.

\subsection{Additional Visualizations}
\label{app:per_task}

\begin{figure}[h]
  \centering
  \includegraphics[width=0.85\linewidth]{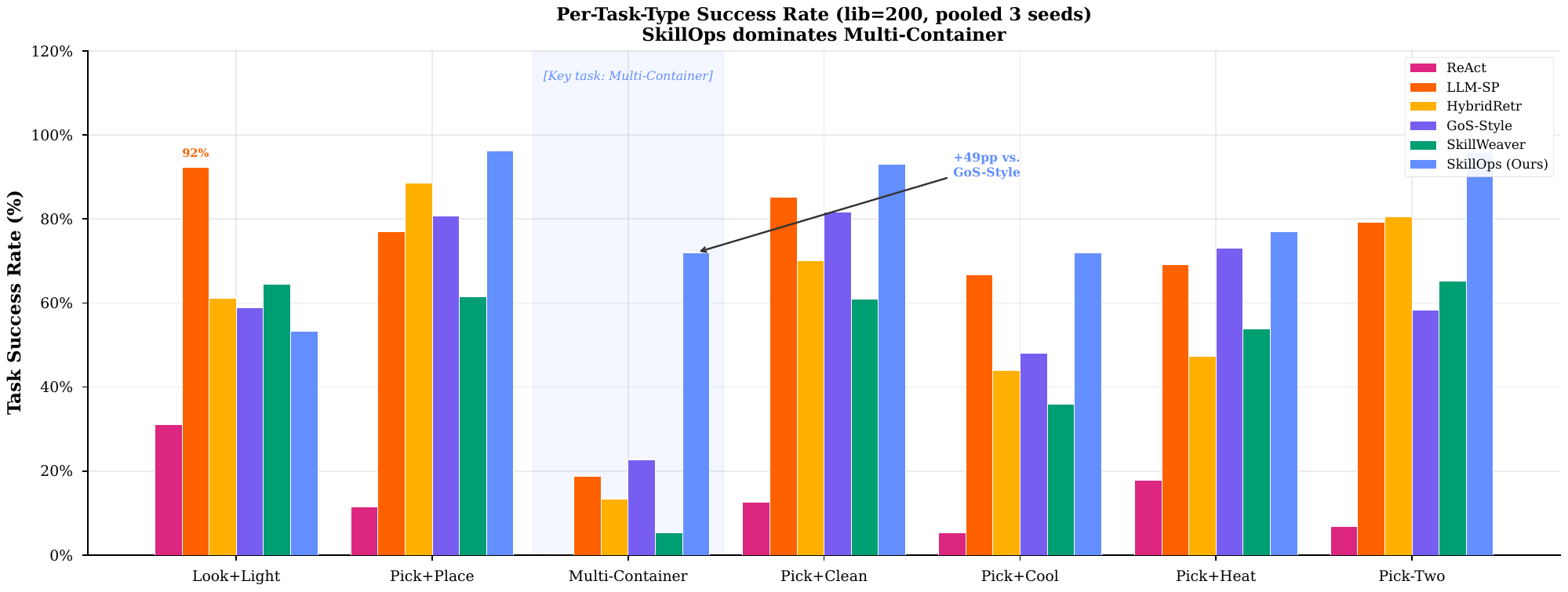}
  \caption{\textbf{Per-task-type SR.} Results are reported for the 200-skill library, pooled over 3 seeds.}
  \label{fig:task_type}
\end{figure}

\begin{figure}[h]
  \centering
  \includegraphics[width=0.7\linewidth]{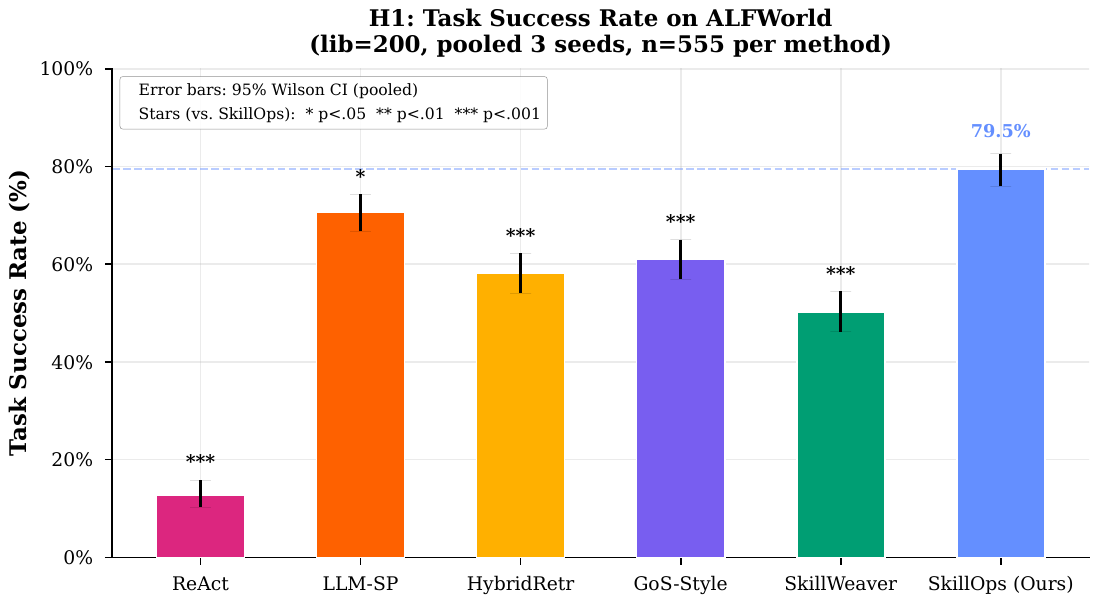}
  \caption{\textbf{H1 main results.} Task SR on ALFWorld at the 200-skill scale, pooled over 3 seeds. Error bars show Wilson 95\% confidence intervals.}
  \label{fig:h1_main}
\end{figure}

\begin{figure}[h]
  \centering
  \includegraphics[width=\linewidth]{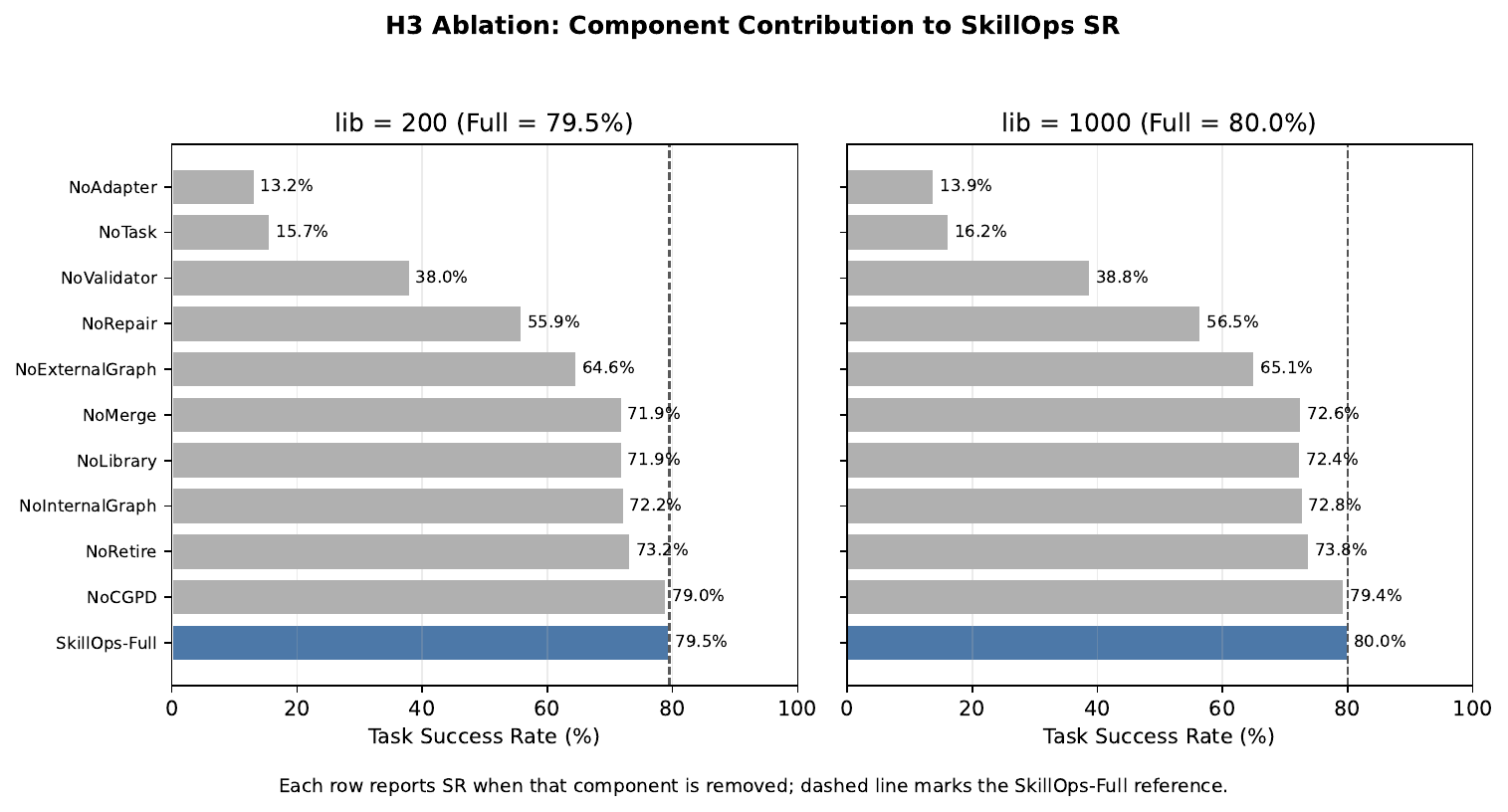}
  \caption{\textbf{H3 ablation visualization.} Each bar reports task SR after removing one SkillOps component.}
  \label{fig:h3_ablation}
\end{figure}

\section{H3 Ablation Study}
\label{app:h3}

Figure~\ref{fig:h3_ablation} visualizes the component-level ablations. The results show that argument binding, task-time planning, and contract structure are central to SkillOps performance, while library-time actions provide additional maintenance capacity for larger or more degraded libraries.


\newpage

%
%

\section{Token Consumption and k-Sensitivity Analysis}
\label{app:tokens}

\subsection{Token Consumption}
\label{app:token_consumption}

We measure the per-task token cost of each method across the nine H2 library scales,
\(N\in\{200,250,500,750,1000,1250,1500,1750,2000\}\), using three random seeds and 185 ALFWorld tasks per seed.
All LLM-based methods use \texttt{temperature=0} and \texttt{gpt-4o-mini}.
The goal is to test whether SkillOps reduces task-time LLM dependence as the library grows.

\begin{table}[h]
\centering
\small
\caption{Total tokens per task (prompt + completion; mean across 3 seeds; $n=185$ trials per cell) at five representative library scales. SkillOps\_Full and SkillOps\_NoMaint emit 0 tokens by typed signature matching (fallback rate = 0\% at all scales). SkillWeaver: 2.93 LLM calls/task flat.}
\label{tab:tokens-summary}
\begin{tabular}{r|ccccc|c}
\toprule
\hline
$N$ & ReAct & LLM-SP & Hybrid & GoS & SkillWeaver & \textbf{SkillOps} \\
\midrule
200  & 359 & 867  & 982  & 636 & 1{,}390 & \textbf{110} \\
500  & 359 & 879  & 985  & 638 & 1{,}387 & \textbf{103} \\
1000 & 359 & 910  & 1{,}011 & 643 & 1{,}395 & \textbf{107} \\
1500 & 359 & 910  & 1{,}014 & 647 & 1{,}397 & \textbf{110} \\
2000 & 359 & 910  & 1{,}017 & 647 & 1{,}390 & \textbf{191} \\
\hline
\bottomrule
\end{tabular}
\end{table}

\begin{figure}[h]
\centering
\includegraphics[width=\linewidth]{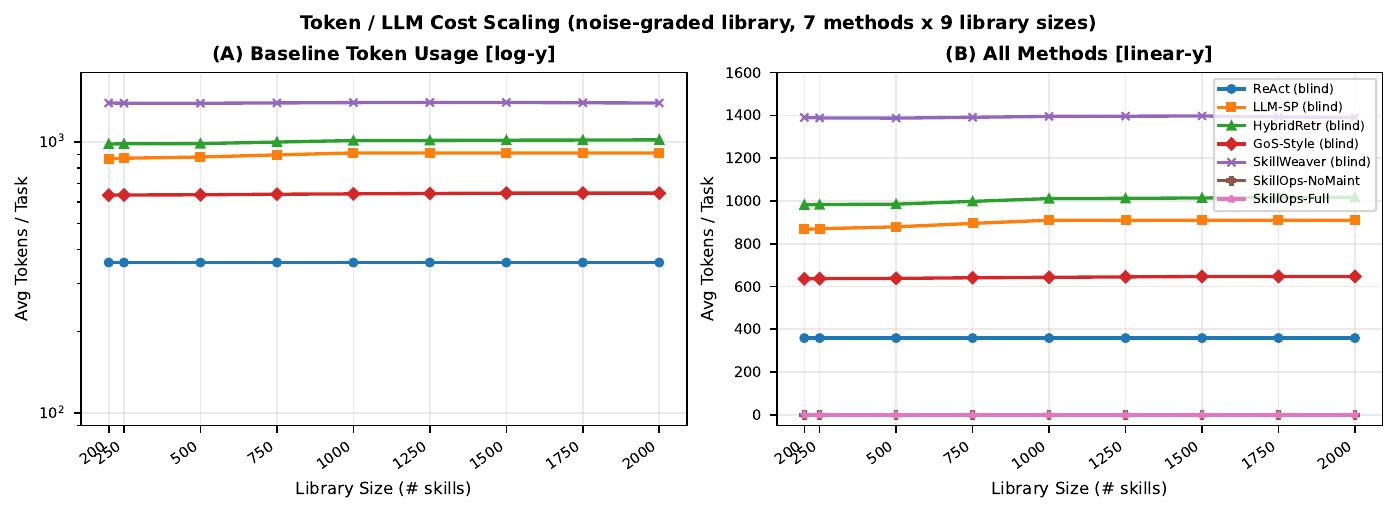}
\caption{\textbf{Token scaling with library size.}
SkillOps emits nearly zero task-time tokens across all evaluated scales, while LLM-based baselines keep nonzero token budgets.}
\label{fig:fig7-tokens}
\end{figure}

Table~\ref{tab:tokens-summary} and Figure~\ref{fig:fig7-tokens} show that SkillOps emits very little LLM tokens per task across all evaluated library sizes. In contrast, LLM-SP, Hybrid, GoS, and SkillWeaver require nonzero prompt and completion tokens because they rely on LLM scoring, planning, or validation at task time. Their token costs remain roughly flat with library size at fixed top-\(k\), since the prompt contains only retrieved skill descriptors rather than the full library. Thus, the main cost advantage of SkillOps is not better asymptotic scaling over retrieval prompts, but the fact that typed contract matching avoids task-time LLM calls in this setting.

\subsection{k-Sensitivity Analysis}
\label{app:k_sens}

We also test whether the gap between SkillOps and retrieval baselines is caused by a small top-\(k\) retrieval budget. We rerun the blind baselines with \(k\in\{6,12,24\}\) at three representative library sizes, \(N\in\{200,1000,2000\}\). SkillOps is \(k\)-free because typed signature matching directly returns a compatible candidate path.

\begin{table}[h]
\centering
\small
\caption{\textbf{k-sensitivity of retrieval baselines.}
SR (\%) is reported for \(k\in\{6,12,24\}\) at three library scales. SkillOps is \(k\)-free and is shown as a fixed reference.}
\label{tab:k_sens}
\begin{tabular}{lrrr|rrr|rrr}
\toprule
\hline
& \multicolumn{3}{c|}{lib=200} & \multicolumn{3}{c|}{lib=1000} & \multicolumn{3}{c}{lib=2000} \\
Method & $k$=6 & $k$=12 & $k$=24 & $k$=6 & $k$=12 & $k$=24 & $k$=6 & $k$=12 & $k$=24 \\
\midrule
\textbf{SkillOps\_Full} & \textbf{79.5} & \textbf{79.5} & \textbf{79.5} & \textbf{80.0} & \textbf{80.0} & \textbf{80.0} & \textbf{80.5} & \textbf{80.5} & \textbf{80.5} \\
\midrule
LLM\_SP\_blind       & 51.0 & 50.8 & 51.9 & 49.7 & 50.8 & 53.5 & 49.4 & 48.1 & 49.2 \\
Hybrid\_blind        & 43.4 & 47.0 & 48.1 & 34.6 & 42.7 & 44.3 & 35.9 & 37.3 & 38.4 \\
GoS\_blind           & 44.1 & 49.2 & 49.2 & 44.0 & 47.0 & 51.9 & 42.0 & 47.0 & 43.8 \\
SkillWeaver\_blind   & 41.6 & 40.0 & 41.1 & 41.4 & 42.2 & 45.4 & 40.9 & 40.5 & 38.9 \\
ReAct\_blind         & 11.4 & 11.4 & 11.4 & 11.9 & 11.4 & 11.4 & 11.9 & 11.4 & 11.4 \\
\hline
\bottomrule
\end{tabular}
\end{table}

\begin{figure}[h]
\centering
\includegraphics[width=\linewidth]{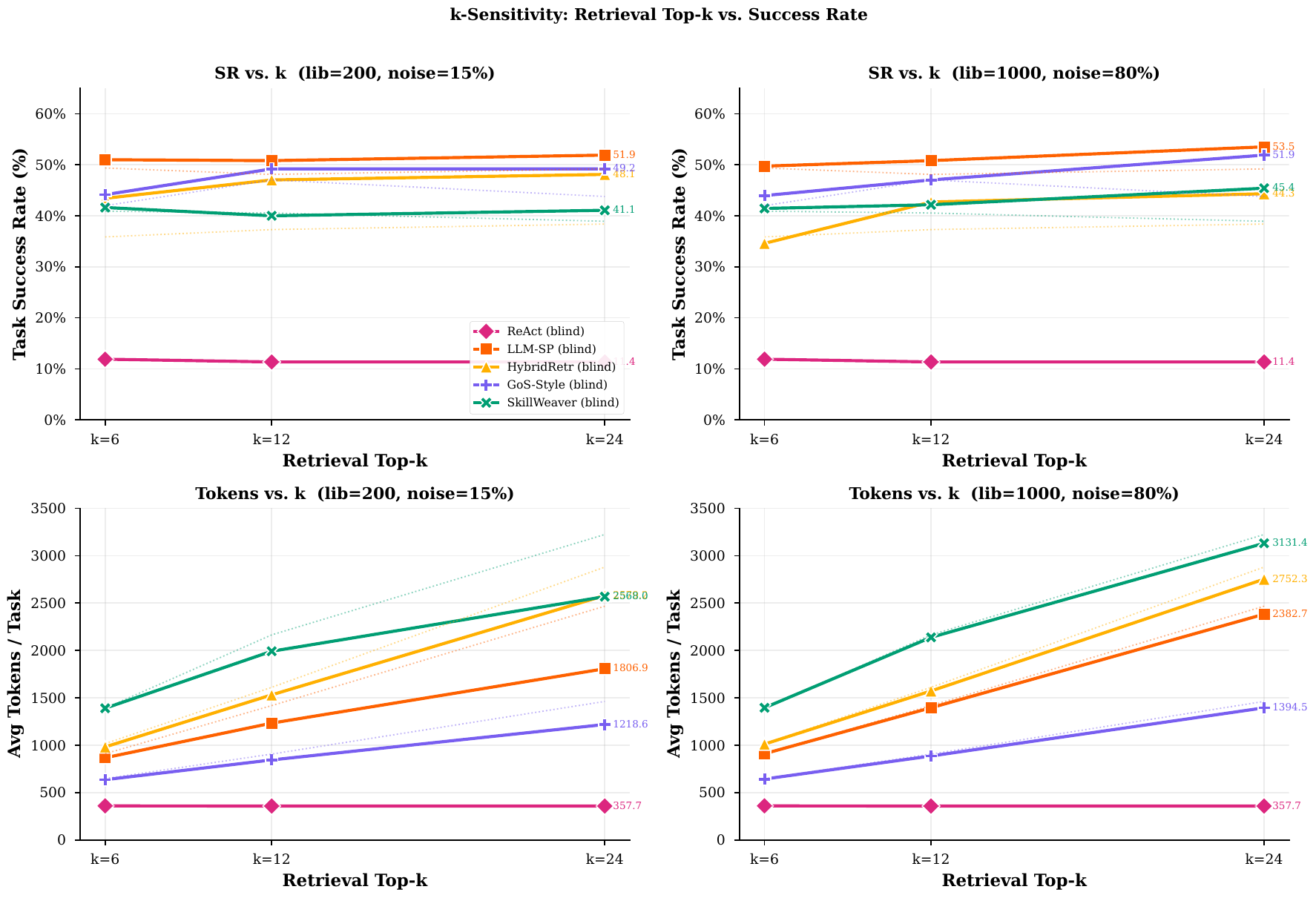}
\caption{\textbf{k-sensitivity visualization.}
Increasing \(k\) improves some retrieval baselines, but SkillOps remains clearly stronger across all tested library sizes.}
\label{fig:fig8-ksens}
\end{figure}

Table~\ref{tab:k_sens} and Figure~\ref{fig:fig8-ksens} show that increasing \(k\) from 6 to 24 improves some baselines, especially Hybrid and GoS, but does not close the gap to SkillOps. At lib=200, the best baseline at \(k=24\) reaches \(51.9\%\) SR, while SkillOps remains at \(79.5\%\). At lib=2000, the best baseline at \(k=24\) reaches \(43.8\%\), while SkillOps remains at \(80.5\%\). This supports the main conclusion that SkillOps's advantage is not simply due to a restrictive retrieval budget for baselines; it comes from typed contract matching and compatibility-constrained planning.

\subsection{Maintenance Overhead: Task-Time, Library-Time, and Amortization}
\label{app:tokens-v5}

We evaluate whether the SkillOps maintenance pass adds hidden overhead during deployment. Specifically, we measure three costs: task-time token change after replacing the raw library with the maintained library, library-time cost for one \texttt{run\_maintenance} pass, and amortized cost per downstream task. The evaluation uses the noise-graded libraries with \(N\in\{200,250,500,750,1000,1250,1500,1750,2000\}\), seven downstream baselines, and three seeds.

\begin{table}[h]
\centering
\small
\caption{\textbf{Task-time token change after maintenance.}
Values report \(\Delta=\text{WithMaint}-\text{NoMaint}\) in percent at five representative library sizes. Negative values indicate fewer task-time tokens after using the maintained library.}
\label{tab:tokens-maint-delta}
\begin{tabular}{r|ccccccc}
\toprule
\hline
$N$ & ReAct & LLM-SP & Hybrid & GoS & SkillWeaver & BM25 & Dense \\
\midrule
200  & \textbf{0.0\%}  & \textbf{$-$0.06\%} & \textbf{$-$0.05\%} & \textbf{$-$0.05\%} & \textbf{$-$0.05\%} & \textbf{+0.28\%} & \textbf{+0.03\%} \\
500  & \textbf{0.0\%}  & $-$1.29\% & \textbf{$-$0.33\%} & $-$1.41\% & $-$3.55\% & \textbf{+5.56\%} & $-$1.62\% \\
1000 & \textbf{0.0\%}  & \textbf{$-$0.91\%} & $-$2.55\% & \textbf{$-$0.96\%} & \textbf{+0.50\%} & +1.43\% & $-$3.95\% \\
1500 & \textbf{0.0\%}  & $-$1.49\% & $-$3.00\% & $-$1.58\% & $-$3.61\% & \textbf{+0.64\%} & $-$2.51\% \\
2000 & \textbf{0.0\%}  & \textbf{+0.03\%} & $-$2.03\% & $-$1.47\% & \textbf{+0.48\%} & $-$1.26\% & $-$2.84\% \\
\hline
\bottomrule
\end{tabular}
\end{table}

Table~\ref{tab:tokens-maint-delta} shows that using the maintained library is usually token-neutral or token-saving at task time. Across the 35 reported baseline-scale cells, 24 cells decrease and 4 are nearly unchanged. The reason is that maintenance prunes or merges low-quality candidates before retrieval, so downstream agents often construct prompts from a cleaner top-\(k\) skill set. This reduces redundant skill descriptions and lowers task-time token usage for most retrieval-based methods.

\begin{table}[h]
\centering
\small
\caption{\textbf{Library-time cost of one maintenance pass.}
The maintained library size, action counts, and estimated LLM overhead are reported at five representative scales. The maintenance pass uses at most 9 LLM calls per library.}
\label{tab:lib-time-cost}
\begin{tabular}{r|cccccc|ccc}
\toprule
\hline
$N$  & before $\to$ after & merge & retire & repair & +valid & +adapt & LLM calls & tokens & cost USD \\
\midrule
200  & 200 $\to$ 185   & 15   & 0  & 5   & 5   & 5   & \textbf{1} & \textbf{1.2K}  & \textbf{\$0.0003} \\
500  & 500 $\to$ 329   & 153  & 18 & 44  & 50  & 44  & \textbf{3} & \textbf{1.6K}  & \textbf{\$0.0006} \\
1000 & 1000 $\to$ 562  & 408  & 30 & 123 & 133 & 118 & \textbf{3} & \textbf{1.0K}  & \textbf{\$0.0003} \\
1500 & 1500 $\to$ 723  & 729  & 48 & 179 & 195 & 171 & \textbf{3} & \textbf{1.4K}  & \textbf{\$0.0003} \\
2000 & 2000 $\to$ 749  & 1209 & 42 & 186 & 200 & 178 & \textbf{4} & \textbf{1.8K} & \textbf{\$0.0003} \\
\hline
\bottomrule
\end{tabular}
\end{table}

Table~\ref{tab:lib-time-cost} shows that the Library-Time Loop adds only a very small maintenance overhead. Most actions are still rule-driven: \texttt{merge} uses body-hash collisions, \texttt{retire} uses utility logs, \texttt{repair} and \texttt{add\_validator} use sibling inheritance when available, and \texttt{add\_adapter} uses type-consistency checks. The limited LLM usage is reserved for compact contract-level edits when needed. Even at the largest scale \(N=2000\), one full maintenance pass uses only 9 LLM calls, about 10.8K tokens, and an estimated cost of \(\$0.0026\). Because this cost is paid once per library update and then amortized over downstream tasks, the deployment overhead remains negligible; the main runtime effect is the task-time token change reported in Table~\ref{tab:tokens-maint-delta}.

\section{ContractGraph-Propagated Diagnosis (CGPD): Algorithm and Results}
\label{app:cgpd}

The recursion above defines a self-map on the space of risk score vectors $\mathbf{R} \in [0,1]^{|\mathcal{S}|}$ under the sup-norm.
Because $\alpha < 1$, this map is a $\alpha$-contraction: each iteration reduces the difference between successive risk estimates by at least a factor of $\alpha$.
By Banach's fixed-point theorem, this guarantees unique convergence from any initialization in at most $O(\log 1/\varepsilon)$ iterations for tolerance $\varepsilon$.
The practical benefit is that CGPD can \emph{preemptively} flag downstream skills for validator insertion before they exhibit failure modes, based solely on the structural topology of the dependency graph.

\subsection{ContractGraph-Propagated Diagnosis (CGPD)}
\label{sec:cgpd}

Skill technical debt can propagate across a skill chain. If an upstream skill produces a malformed artifact, a downstream skill may fail even when its own operation is correct. Diagnosing each skill independently can therefore miss cascading failures. To address this issue, \textbf{ContractGraph-Propagated Diagnosis (CGPD)} propagates risk scores along dependency edges in HSEG and identifies downstream skills that should receive preventive maintenance, such as validators or adapters.

\subsubsection{Risk Propagation Model}

Let the ecosystem graph be \(\mathcal{G}=(\mathcal{S},\mathcal{R})\). For each skill \(s\in\mathcal{S}\), let \(R_{\mathrm{loc}}(s)\in[0,1]\) denote its local risk score, computed from observable health signals such as missing validators, high failure rate, interface mismatch, or low utility. CGPD converts these local scores into propagated risk scores by passing risk along dependency edges. Let
\[
\mathrm{Parents}(s)=\{s'\in\mathcal{S}:s'\xrightarrow{\mathrm{dep}}s\in\mathcal{R}\}
\]
denote the upstream skills whose artifacts are consumed by \(s\). Starting from \(R^{(0)}(s)=R_{\mathrm{loc}}(s)\), CGPD updates
\begin{equation}
R^{(t+1)}(s)
=
(1-\alpha)R_{\mathrm{loc}}(s)
+
\alpha
\max_{s'\in \mathrm{Parents}(s)}R^{(t)}(s'),
\label{eq:risk_prop}
\end{equation}
where \(\alpha\in(0,1)\) controls the amount of upstream risk propagated to the current skill. The max operator reflects a worst-upstream-risk rule: a downstream skill can become unsafe if any one of its required upstream artifacts is unreliable.

\subsubsection{Convergence Property}

CGPD is training-free and does not require labeled failure data. The update in Equation~\ref{eq:risk_prop} defines a contraction under the \(\ell_\infty\) norm because the upstream term is weighted by \(\alpha<1\). Therefore, by the Banach fixed-point theorem~\citep{banach1922operations}, the iteration converges to a unique fixed point \(R^{(\infty)}\). In acyclic dependency graphs, the propagated scores can also be computed in a finite number of passes bounded by the graph depth.

\begin{algorithm}[ht]
\caption{ContractGraph-Propagated Diagnosis (CGPD)}
\label{alg:cgpd}
\begin{algorithmic}[1]
\Require Ecosystem graph \(\mathcal{G}=(\mathcal{S},\mathcal{R})\), local risk \(R_{\mathrm{loc}}\), propagation weight \(\alpha\in(0,1)\), threshold \(\tau\)
\Ensure Maintenance trigger set \(\mathcal{T}\)
\State \(R^{(0)}(s)\gets R_{\mathrm{loc}}(s)\) for all \(s\in\mathcal{S}\)
\For{\(t=0,\ldots,T-1\)}
    \For{each skill \(s\in\mathcal{S}\)}
        \State \(R_{\mathrm{in}}(s)\gets \max_{s'\in\mathrm{Parents}(s)}R^{(t)}(s')\)
        \State \(R^{(t+1)}(s)\gets (1-\alpha)R_{\mathrm{loc}}(s)+\alpha R_{\mathrm{in}}(s)\)
    \EndFor
    \If{\(\max_{s\in\mathcal{S}}|R^{(t+1)}(s)-R^{(t)}(s)|<\varepsilon\)}
        \State \textbf{break}
    \EndIf
\EndFor
\State \(\mathcal{T}\gets \mathrm{ApplyBasicHeuristics}(\mathcal{G})\)
\For{each skill \(s\in\mathcal{S}\)}
    \If{\(R^{(t+1)}(s)>\tau\) and \(V_s=\emptyset\)}
        \State \(\mathcal{T}\gets \mathcal{T}\cup\{\texttt{add\_validator}(s)\}\)
    \EndIf
\EndFor
\State \Return \(\mathcal{T}\)
\end{algorithmic}
\end{algorithm}

Algorithm~\ref{alg:cgpd} first initializes each skill with its local risk, then iteratively propagates upstream risk through dependency edges until convergence. After convergence, CGPD flags high-risk skills without validators for \texttt{add\_validator}. This makes the maintenance decision graph-aware: a skill can be selected for preventive maintenance not only because it is locally risky, but also because it inherits risk from upstream dependencies.

\subsection{CGPD Results}

\begin{table}[h]
\centering
\small
\caption{\textbf{CGPD results.}
We compare basic SkillOps maintenance with CGPD-augmented maintenance at two large library sizes. Results are averaged over 3 seeds.}
\label{tab:cgpd_mve}
\begin{tabular}{lcc}
\toprule
\hline
Variant & lib$=$1000 SR & lib$=$2000 SR \\
\midrule
SkillOps\_Full (basic) & $79.5\%\pm0.0$pp & $80.5\%\pm0.0$pp \\
SkillOps\_Full\_CGPD & $\mathbf{80.0\%\pm0.0}$pp & $\mathbf{81.1\%\pm0.0}$pp \\
\midrule
CGPD $-$ basic & $\mathbf{+0.5}$pp & $\mathbf{+0.6}$pp \\
\hline
\bottomrule
\end{tabular}
\end{table}

Table~\ref{tab:cgpd_mve} shows that adding CGPD improves SkillOps over the basic maintenance variant at both large library sizes. The gains are modest but consistent: \(+0.5\)pp at lib\(=1000\) and \(+0.6\)pp at lib\(=2000\). This supports the role of dependency-aware risk propagation: beyond local health rules, CGPD can identify downstream risks induced by upstream skills and trigger preventive maintenance before those risks appear as task failures. The improvement is larger at the higher-noise scale, suggesting that graph-based risk propagation becomes more useful as skill technical debt accumulates.


\end{document}